%% file: HighTempFG_with-all-tex.tex
\documentclass[aps,pra,twocolumn,superscriptaddress]{revtex4}
\usepackage{amsmath}
\DeclareMathOperator{\Tr}{Tr}
\usepackage{amsfonts}
\usepackage{graphicx}
\usepackage{float}

\usepackage{tikz}
\usetikzlibrary{calc,decorations.markings,snakes}
\usetikzlibrary{arrows,decorations.pathmorphing,backgrounds,positioning,fit,petri}
\tikzset{
	every picture/.style={font issue={\fontsize{6}{12}}, font=\bfseries ,line width =0.8pt},
         font issue/.style={execute at begin picture={#1\selectfont}},
     	T2/.style={draw=black,decorate, 
			decoration={border, segment length=2pt,amplitude=1pt, angle=0}},
        }

\begin{document}

\title{High-temperature limit of the resonant Fermi gas}

\author{Vudtiwat Ngampruetikorn}
\affiliation{T.C.M. Group, Cavendish Laboratory, JJ Thomson Avenue, Cambridge,
 CB3 0HE, United Kingdom} %

\author{Meera M. Parish}
\affiliation{London Centre for Nanotechnology, Gordon Street, London, WC1H 0AH, United Kingdom}

\author{Jesper Levinsen}
\affiliation{Aarhus Institute of Advanced Studies, Aarhus University, DK-8000 Aarhus C, Denmark}

\date{\today}

\newcommand{\ve}[1]{{\mathbf #1}}
\newcommand{\up}{\uparrow}
\newcommand{\down}{\downarrow}
\renewcommand{\k}{{\bf k}}
\newcommand{\eb}{\varepsilon_\text{b}}
\newcommand{\p}{{\bf p}}
\newcommand{\q}{{\bf q}}
\newcommand{\0}{{\bf 0}}
\newcommand{\x}{\hat {\bf x}}
\newcommand{\y}{\hat {\bf y}}
\newcommand{\z}{\hat {\bf z}}
\newcommand{\R}{\bf R}
\renewcommand{\r}{{\bf r}}
\newcommand{\bra}[1]{\langle\left.{#1}\right|}
\newcommand{\ket}[1]{\left|{#1}\right.\rangle}
\newcommand{\ef}{\varepsilon_F}
\newcommand{\eq}{\epsilon_{\q}}
\newcommand{\eqcm}{\epsilon_{\q,\text{cm}}}
\newcommand{\ep}{\epsilon_{\p}}
\newcommand{\ek}{\epsilon_{\k}}
\newcommand{\ekq}{\epsilon_{\q-\k}}
\newcommand{\ekkq}{\epsilon_{\k+\k'-\q}}
\newcommand{\nn}{\nonumber}
\newcommand{\op}{\omega_z}
\newcommand{\T}{{\cal T}}
\newcommand{\F}{{\cal F}}
\newcommand{\n}{\vec n}
\newcommand{\e}[1]{\epsilon_{#1}}
\renewcommand{\O}{\Omega}
\renewcommand{\P}{\ve P}

\newcommand{\jesper}[1]{{\color{blue}#1}}
\newcommand{\meera}[1]{{\color{red} #1}}

\begin{abstract}
  We use the virial expansion to investigate the behavior of the
  two-component, attractive Fermi gas in the high-temperature limit,
  where the system smoothly evolves from weakly attractive fermions to
  weakly repulsive bosonic dimers as the short-range attraction is
  increased.
We present a new formalism for computing the virial coefficients that
employs a diagrammatic approach to the grand potential and allows one
to easily include an effective range $R^*$ in the interaction. In the
limit where the thermal wavelength $\lambda \ll R^*$, the calculation
of the virial coefficients is perturbative even at unitarity and the
system resembles a weakly interacting Bose-Fermi mixture for all
scattering lengths $a$. By interpolating from the perturbative limits
$\lambda/|a| \gg 1$ and $R^*/\lambda \gg 1$, we estimate the value of
the fourth virial coefficient at unitarity for $R^*=0$ and we find
that it is close to the value obtained in recent experiments. We also
derive the equations of state for the pressure, density and entropy,
as well as the spectral function at high temperatures.
\end{abstract}

\maketitle


\section{Introduction}

It was long ago argued that the separate phenomena of BCS pairing of
fermions, observed in conventional superconductors, and Bose-Einstein
condensation (BEC) can be smoothly
connected~\cite{Eagles1969,Leggett1980,Comte1982}. Such a BCS-BEC
crossover may be realized by increasing the strength of attractive
interactions between spin-$\up$ and $\down$ fermions, 
so that the system evolves from weakly bound $\up$-$\down$ pairs in
the BCS limit to strongly bound bosonic dimers in the BEC regime. 
This scenario has been successfully achieved in recent experiments on 
ultracold atomic Fermi
gases~\cite{Regal2004,Zwierlein2004,chin2004,bourdel2004,kinast2004,zwierlein2005}.
In the low-temperature superfluid phase, the same $U(1)$ symmetry is
broken in the BCS and BEC limits, and for $s$-wave pairing one expects
a smooth crossover between these limits without any phase transition. 
However, the nature of the low-energy quasiparticle excitations
changes markedly as the interactions are varied and this affects the
behavior of the system above the superfluid transition temperature
$T_c$: For weak attraction, the normal state at low temperatures
corresponds to a Fermi liquid, while for strong attraction, one
instead obtains a weakly repulsive Bose liquid just above
$T_c$~\cite{Nozieres1985}. 
Thus, the BCS-BEC crossover at zero temperature 
is associated with a crossover from fermionic to bosonic behavior in
the normal state.

The different character of the two limits has been dramatically
revealed in spin-imbalanced Fermi gases.
At zero temperature,
if one has a large difference in the spin 
densities $n_\sigma$, i.e., $n_\up \gg n_\down$, 
then one obtains a first-order phase transition 
between a Fermi-liquid phase and a superfluid phase composed of a
Bose-Fermi mixture~\cite{sheehy2007,Parish2007,pilati2008}. 
Moreover, in the limit of extreme imbalance where $n_\down \to 0$, it
was shown that the ground state of a single spin-down particle
undergoes an abrupt transition from a fermionic to bosonic
quasiparticle as the attraction is
increased~\cite{prokofiev2008,punk2009,mora2009,combescot2009}. 
However, for the unpolarized case, the manner in which fermions evolve
into bosons at finite temperature is not well understood, 
in part because of the difficulty in theoretically treating the intermediate regime of attraction between the BCS and BEC limits.
Here, we elucidate the Fermi-Bose crossover of the unpolarized gas in the \emph{high-temperature} limit, where one can perform a controlled calculation by exploiting the virial expansion.

\begin{figure}
\centering
\includegraphics[width=0.9\linewidth]{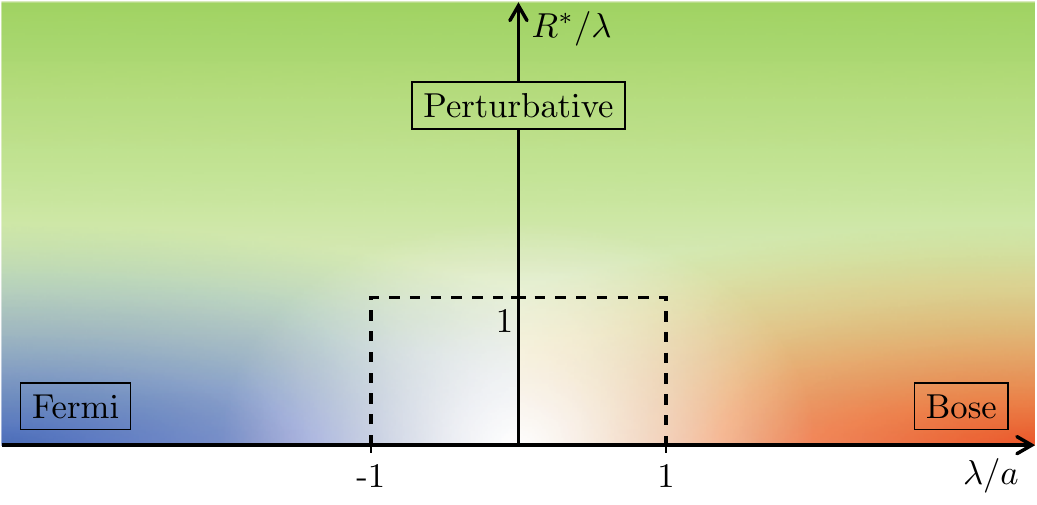}
\caption{Different interaction regimes for the high-temperature 
  unpolarized Fermi gas at density $n\ll \lambda^{-3}$.  When the scattering
  length $a <0$ and the thermal de Broglie wavelength $\lambda \gg
  |a|$, the system corresponds to a weakly attractive Fermi gas. In
  the opposite limit, where $\lambda \gg a >0$, the system instead
  resembles a weakly repulsive gas of bosonic dimers once 
  the two-body binding energy $\eb \gg k_B T \log(\sqrt{2}/n\lambda^3)$,
 where $T$ is the temperature.
  In the crossover region $\lambda/|a| \lesssim1$, the system is strongly
  interacting in the sense that each coefficient of the virial
  expansion cannot be obtained from a perturbative calculation around
  a non-interacting system.  This region can be made perturbative by
  introducing a large effective range parameter $R^* > \lambda$.}
\label{fig:regimes}
\end{figure}

The high-temperature limit corresponds to a gas far from quantum
degeneracy, where the thermal wavelength $\lambda \ll k_F^{-1}$, with
the Fermi wavevector $k_F = (3\pi^2 n)^{1/3}$ 
and density $n = n_\up +n_\down$. 
 Assuming
short-range interactions characterized by scattering length $a$, the
crossover in this limit is thus parametrized by the dimensionless
ratio $\lambda/a$, with the system becoming a gas of bosonic dimers
once $\lambda/a$ is sufficiently large 
(Fig.~\ref{fig:regimes}).  This is to be
contrasted with the BCS-BEC crossover at zero temperature, which is
instead dependent on $1/k_F a$. 
Here, the crossover from fermionic to bosonic behavior may be defined
as the point where the chemical potential $\mu$ crosses zero, since
this is where BCS mean-field theory predicts a qualitative change in
the quasiparticle dispersion~\cite{Leggett1980,Parish2005}. Indeed, a
very recent experiment performed 
at temperatures $T \gtrsim T_c$ has observed the
breakdown of Fermi-liquid behavior 
at an interaction $1/k_Fa$ close to this point~\cite{Sagi2014}. 
The low-temperature crossover is qualitatively different from its
high-temperature counterpart since it is connected with the existence
of a Fermi surface rather than with the thermal dissociation of
dimers.  However, both temperature regimes are strongly interacting
near unitarity $1/a=0$, since one cannot perform a perturbative
expansion in $a$ around a non-interacting system, 
unlike in the opposite limit $|a| \to 0$. 

The virial or cluster expansion is one way of perturbing around a
non-interacting limit for arbitrary $a$. In this case, the
non-interacting system is the classical Boltzmann gas and the
expansion parameter is the fugacity $z=e^{\beta \mu}$, where
$\beta^{-1} = k_B T$ 
and $k_B$ is the Boltzmann constant (which we set to 1 in the following). 
The coefficients of the different powers of $z$ in the virial
expansion are then functions of $\lambda/a$ and determine the
crossover from fermionic atoms to bosonic dimers for a given value of
$z$. While the second order virial coefficient for short-range interactions
was determined long ago~\cite{BethUhlenbeckII}, despite early
attempts at extending the expansion -- see, e.g., Ref.~\cite{Pais1959} --
the extension of the virial expansion to higher powers in fugacity has
only recently been achieved, mainly motivated by high precision
experiments on ultracold Fermi gases. In particular, at unitarity,
virial coefficients for the expansion up to $z^4$ have been determined
both theoretically~\cite{Liu2009,Leyronas2011,Kaplan2011,Rakshit2012}
and experimentally~\cite{Salomon2010,Zwierlein2012}. However, the
complexity of the problem increases exponentially with the order of
the coefficient, and indeed it is already challenging to accurately
compute the fourth virial coefficient, as exemplified by the current
discrepancy between the theoretical prediction \cite{Rakshit2012} and
experiment \cite{Salomon2010,Zwierlein2012} at unitarity. For a recent
review of the virial expansion as applied to the strongly interacting
Fermi gas we refer the reader to Ref.~\cite{LiuReview}.

One route to simplifying the calculation of the virial coefficients is
to consider an interaction with an effective range, characterized by
the additional length scale $R^*$~\footnote{
The effective range occurring for zero-range interactions 
should, of course, be distinguished from a finite range interaction, such as 
used in Ref.~\cite{Kokkelmans2002}.
}. In the limit of a ``narrow
resonance'' $R^*/\lambda \gg 1$, we can then perform a perturbative
expansion in $\lambda/R^*$ for any $a$, as depicted in
Fig.~\ref{fig:regimes}. This is analogous to the situation at zero
temperature, where the calculation for the BCS-BEC crossover is
perturbative when $k_F R^* \gg 1$~\cite{Gurarie2007}. In contrast to
Ref.~\cite{Ho2012}, we regard the narrow-resonance case as
\emph{weakly interacting} since the limit $R^* \to \infty$ corresponds
to a non-interacting mixture of fermionic atoms and bosonic dimers, as
discussed below. Note that such a weakly interacting system can still
exhibit a large interaction energy (e.g., at unitarity) owing to the
contribution of the binding energy of the dimers.

In this paper, we present the virial expansion of the resonant Fermi gas
throughout the Fermi-Bose crossover for various $R^*/\lambda$. 
To this end, we develop a new formalism for computing virial
coefficients that employs a diagrammatic approach to the grand
potential similar to that in Ref.~\cite{Dashen1969}, but where we use
a $T$-matrix rather than an $S$-matrix description for the scattering
of particles. We show that our expression for the second virial
coefficient is equivalent to the well-known Beth-Uhlenbeck formula
involving scattering phase shifts \cite{BethUhlenbeckII}.  
We also, for the first time, extend the virial expansion to higher powers of $z$
 for finite $R^*$.
By considering the perturbative limits $\lambda/|a| \gg 1$ and
$R^*/\lambda \gg 1$, we extrapolate our results and obtain an approximate value for the fourth
virial coefficient at unitarity and $R^*=0$. We also determine the
equation of state in the high-temperature limit and compare with the
recent MIT experiment \cite{Zwierlein2012}.

The paper is organized as follows. In Sec.~\ref{sec:FG} we
introduce the resonant Fermi gas as described by a two-channel model,
and we discuss the role of the parameters of the model. In Sec.\
\ref{sec:virial} we present the formalism for the virial expansion; for generality, 
we consider any possible spin- and mass-imbalance, any dimensionality,
and any short-range model of the interactions. 
Section \ref{sec:vcs}
then concerns the virial expansion as applied to the resonant Fermi
gas; we discuss 
the high-temperature crossover from the
Fermi to Bose regime, particularly the limiting behavior of the
virial coefficients, as well as how the system evolves towards a
non-interacting system upon increasing the ratio of the effective
range to thermal wavelength. We present exact calculations of the
second and third virial coefficients, and a perturbatively exact
calculation of the fourth. In Sec.~\ref{sec:thermodyn} we display
the high-temperature behavior of thermodynamic quantities, i.e.,
pressure, density, and entropy. In Sec.~\ref{sec:RF} we discuss the
momentum-resolved radio-frequency spectrum which one would measure in the resonant Fermi
gas, focusing on the difference between broad and narrow
resonances. We conclude in Sec.~\ref{sec:conc}.


\section{Resonant Fermi Gas \label{sec:FG}}

To describe the resonant Fermi gas, we use a specific model relevant 
to ultracold atomic gases with short-range $s$-wave interactions, 
namely the two-channel model, introduced for bosons in Ref.~\cite{Drummond1998}. 
This model describes how the interaction between two atoms scattering freely 
(in the open channel) can become resonant when they are coupled to 
a molecular state (closed channel) and the energy of the collision is 
close to that of the molecule. For the two-component Fermi gas, 
the Hamiltonian is
\begin{align}
  \hat H = \sum_{\k\sigma}\epsilon_\k\hat c^\dag_{\k\sigma}\hat
  c_{\k\sigma} + \sum_{\p}\left(\omega_0+\frac{p^2}{4m}\right)\hat
  b^\dag_{\p}\hat b_{\p} \nn\\+\frac{g}{\sqrt
    V}\sum_{\k\p}\left(b^\dag_{\p}\hat c_{\frac{\p}{2}+\k,\up}\hat
    c_{\frac{\p}{2}-\k,\down} +\text{h.c.}\right),
  \label{eq:2ch}
\end{align}
where the operator $\hat c_{\k \sigma}^\dag$ ($\hat c_{\k \sigma}$)
creates (annihilates) a fermionic atom with spin $\sigma$, momentum
$\k$, mass $m$ and single-particle energy $\ek=k^2/2m$; $\hat b_{\p}^\dag$
($\hat b_{\p}$) creates (annihilates) a closed-channel molecule with
momentum $\p$. $g$ denotes the strength of the coupling which converts
a pair of atoms into a closed-channel molecule --- we take this to be 
constant up to an ultraviolet momentum cutoff $\Lambda$. $\omega_0$ is
the bare detuning, $V$ is the volume, and we set 
$\hbar =1$.

Following renormalization of the contact interaction, this Hamiltonian
leads to the two-body $T$ matrix
\begin{align}\label{2body-Tmat}
T_2(E) =
\frac{4\pi/m}{a^{-1}-\sqrt{-mE}+mR^*E},
\end{align}
where the bare detuning $\omega_0$ is parameterised by the scattering
length $a=(\frac{2\Lambda}{\pi}-\frac{4\pi\omega_0}{mg^2})^{-1}$ and
range parameter $R^*=\frac{4\pi}{m^2g^2}$. For positive scattering
length, there exists a two-body bound state (dimer) with binding energy,
\begin{align}
\eb=\frac{\left[\sqrt{1+4R^*/a}-1\right]^2}{4mR^{*2}}.
\end{align}
This corresponds to a pole of the $T$ matrix with residue
\begin{align}\label{eq:residue}
g^2Z=\frac{4\pi}{m^2R^*}\left(1-\frac{1}{\sqrt{1+4R^*/a}}\right).
\end{align}

The gas in the resonant regime is characterized by a scattering length
$a$ which greatly exceeds the van der Waals range of the interactions,
$R_e$, and we shall assume this to be the case in the
following. Indeed, near a magnetic field Feshbach resonance, the
scattering length behaves as
\begin{align}
a=a_{\rm bg}\left[1-\frac{\Delta B}{B-B_0}\right],
\end{align}
and can be much larger than the off-resonant background scattering length
$a_{\rm \tiny bg}$ which is typically of the order of the van der
Waals range. Here $B$ is the magnetic field, while $B_0$ and $\Delta
B$ are the location and magnetic field width of the Feshbach
resonance. The effective range $r_0$, on the other hand, remains
nearly constant across the resonance. It is typically set by the van
der Waals range, but it is large and negative for Feshbach resonances
which are narrow in magnetic field width such that $\mu_{\rm\tiny
  rel}|\Delta B|\ll 1/mR_e^2$ with $\mu_{\rm \tiny rel}$ the
difference in magnetic moment of the two channels.  Then it is
convenient to define instead~\cite{Petrov2004}
\begin{align}
R^*=-r_0/2=\frac1{ma_{\rm \tiny bg}\mu_{\rm \tiny rel}\Delta B},
\end{align}
which relates the effective range to the experimental
parameters. Thus, in the following we can use the two-body $T$ matrix
\eqref{2body-Tmat} to describe a realistic Feshbach resonance where
the scattering length $a$ can diverge, while the range parameter $R^*$
is either negligible or positive.

At finite temperature, where we have thermal wavelength $\lambda = \sqrt{2\pi/mT}$,
the spin- and mass-balanced Fermi gas described
by the Hamiltonian \eqref{eq:2ch} contains three dimensionless
parameters: 
\begin{align}
z, \hspace{10mm} \lambda/a, \hspace{10mm} \lambda/R^*.
\end{align}
The fugacity controls the accuracy of the virial expansion, 
while the second parameter $\lambda/a$ determines the crossover from a weakly attractive Fermi gas to a
weakly repulsive Bose gas. 
In order to have well-defined dimers, we clearly require $\eb \gg T$ or equivalently  $\lambda/a \gg 1$.
However, 
entropy also plays a role at high temperature such that we always 
recover a Boltzmann gas of unbound atoms for sufficiently small $z$, i.e., for $|\mu| \gg \eb$. Thus, we must also consider 
the chemical potential (or density) of the gas when determining the regime where we have a gas of dimers.  As shown in the following sections, the density in the limit $z \ll 1$ and $\eb \gg T$ is given by
\begin{align}
n \simeq \frac{2}{\lambda^3} \left(z + 2\sqrt{2} e^{\beta\eb} z^2 + \ldots 
\right).
\end{align}
The second term effectively corresponds to a Boltzmann gas of dimers (where $\mu_{\rm Bose} = 2\mu + \eb$) while the first term is the usual Boltzmann expression for unbound atoms. Thus, we obtain the bosonic dimer regime when the second term dominates over the first, i.e., 
$2\sqrt{2} e^{\beta\eb} z \gg 1$. This yields the condition $n \lambda^3 \gg \sqrt{2} e^{-\beta\eb}$, which is equivalent to the expression in the caption of Fig.~\ref{fig:regimes}.
Note, further, that when the majority of particles are bound into dimers,
one obtains $\mu \lesssim -\eb/2$. 
Thus, $z \ll 1$ always, which naively suggests that the virial expansion holds all the way down to $T=0$. 
However, this instead means that we must modify the virial expansion to reflect the change in quasiparticles so that 
the relevant expansion parameter is now
$z^* = e^{\beta\eb/2} z$ --- see also Ref.~\cite{Dashen1969} and Sec.~\ref{sec:vcs}.

The third dimensionless parameter,
$\lambda/R^*$, 
in turn, allows a perturbative expansion in small $\lambda/R^*$ of the virial
coefficients themselves.  
This is a consequence of the typical collision energies, set by the
temperature, greatly exceeding the interaction energy scale $\sim
g^4\propto1/R^{*2}$, rendering the system effectively perturbative in
the bare coupling constant. This behavior mimics that of the many-body
problem at zero temperature, which is perturbative for all scattering
lengths if $R^*$ greatly exceeds the average interparticle spacing
\cite{Gurarie2007}. Indeed, in that work a
narrow Feshbach resonance was defined as one fulfilling $k_FR^*\gg1$,
and here we extend this definition to mean any resonant Fermi gas for
which the typical interaction energies greatly exceed that set by the
bare coupling. Thus, in the high-temperature limit of the resonant
Fermi gas, the broad (narrow) resonance limits correspond to
$\lambda/R^*\gg1$ ($\lambda/R^*\ll1$), respectively.


\section{Virial expansion \label{sec:virial}}

We now present the virial expansion of the grand potential for a
two-component ($\up$, $\down$) Fermi system with contact interspecies
interaction. For the sake of generality, we do not restrict ourselves
to a system described by the Hamiltonian \eqref{eq:2ch}; instead we
consider the problem in $d$ dimensions and for arbitrary mass ratio
$m_\up/m_\down$. We note that the formalism outlined in this section
may be straightforwardly extended to more fermionic components, a Bose
gas, or even a Bose-Fermi mixture. In Sec.\ \ref{sec:vcs} we apply
the results of the present section to the resonant Fermi gas described
by the two-channel Hamiltonian \eqref{eq:2ch}.

For a Hamiltonian $\hat H$, the grand potential is given by (see,
e.g., \cite{AGD})
\begin{align}
-\beta\O
&= \sum_{\{N_\sigma\}} \left[
\Tr_{\{N_\sigma\}} \hat A
e^{-\beta(\hat H-\sum_\sigma\mu_\sigma\hat N_\sigma)}\right]_c
\nn \\
\label{eq:OmegaDef}
&= \sum_{N_\up N_\down} z_\up^{N_\up}z_\down^{N_\down} 
\left[\Tr_{N_\up N_\down} \hat A e^{-\beta\hat H}\right]_c,
\end{align}
where $\hat N_\sigma$ is the number
operator of $\sigma$ particles, $\mu_\sigma$ is the chemical potential
for each spin,
and $z_\sigma = e^{\beta\mu_\sigma}$.  The suffix $c$ indicates that
only connected diagrams are kept and the operator $\hat A$ represents
the sum of all identical particle permutations with negative signs
affixed in front of odd permutations: $\hat A\equiv\sum_P(-1)^PP$
where $P$ is a permutation acting only among identical particles, with
$(-1)^P$ the sign of the permutation. The sum over non-negative
$N_\sigma$ is such
that the total number of particles $\sum_\sigma N_\sigma$ is 
positive. The trace on states of the system is taken at fixed particle
number. 
When the temperature is high or the system dilute, the thermodynamics
of the system can be accurately described by the first few terms of
the above power series. That is, the thermodynamics is essentially determined by
few-body physics. However, to establish such a connection, it is
convenient to separate the dynamical (microscopic) and statistical
information \cite{Dashen1969}.

As a starting point, we define the Green's operators 
(see also Ref.~\cite{Dashen1969})
\begin{align}
\hat G(E) = \frac{1}{E-\hat H}, \quad \hat G_0(E) = \frac{1}{E-\hat H_0},
 \end{align}
such that 
\begin{align}\label{eq:thermaloperator_old}
 e^{-\beta\hat H} = 
\oint \frac{dE}{2\pi i} e^{-\beta E} \hat G(E).
\end{align}
This formulation exactly achieves the goal of separating the
statistical information from the dynamical, since $\hat G$ does not
depend on temperature. Here, the Hamiltonian $\hat H=\hat H_0+\hat
H_\text{int}$ consists of the non-interacting part $\hat H_0$ and the
interactions $\hat H_\text{int}$. $E$ is a complex variable and the
counterclockwise contour encloses the structure of the Green's
operator on the real axis, i.e., the integral picks out the spectrum
of $\hat H$. [We let $\oint$ denote a counterclockwise contour
throughout this work.]  For brevity, we define the linear operator
\begin{align}
\oint'_E \equiv \oint \frac{dE}{2\pi i} e^{-\beta E},
\end{align}
such that Eq.~(\ref{eq:thermaloperator_old}) reads
\begin{align}\label{eq:thermaloperator}
 e^{-\beta\hat H} = \oint'_E \hat G(E).
\end{align}
The operator $\oint'_E$ is the Laplace transform 
which connects the few-body dynamics to the statistical
physics. Substituting Eq.~(\ref{eq:thermaloperator}) in
Eq.~(\ref{eq:OmegaDef}), we obtain
\begin{align}
\label{eq:grandpotential}
-\beta \O 
= \sum_{N_\up N_\down} z_\up^{N_\up}z_\down^{N_\down} 
\oint'_E
\left[\Tr_{N_\up N_\down} \hat A \hat G(E)\right]_c .
\end{align}
Physically, the imaginary part of $[\dots]_c$ on the real axis ($\Im
E\rightarrow0^+$) is the spectral density of the few-body system and
the contour integral with a Boltzmann factor is therefore equivalent
to counting all accessible states in the system.

Note that one must be careful when dealing with processes which 
contain identical subclusters: If a diagram has $M$ distinct
sub-clusters, the $i$'th of these appearing $c_i$ times, then a factor of
$(c_1!c_2!\cdots c_M!)^{-1}$ must be introduced when applying $\hat A$
to avoid double counting. This will become clear when we consider the
non-interacting contributions in Section \ref{sect:Bfree}.

We define the virial coefficients, $B_{N_\up N_\down}$, such that
\begin{align}\label{eq:B_definition}
-\beta\O = \frac{V}{\lambda_r^d}\sum_{N_\up N_\down}B_{N_\up N_\down}
z_\up^{N_\up} z_\down^{N_\down}.
\end{align}
where the `reduced' thermal wavelength is given by $\lambda_r \equiv
\sqrt{\pi/m_rT}$, with the reduced mass
$m_r^{-1}=m_\up^{-1}+m_\down^{-1}$.  From
Eqs.~(\ref{eq:grandpotential}) and (\ref{eq:B_definition}) we identify
\begin{align}
B_{N_\up N_\down}
= \frac{\lambda_r^d}{V} \oint'_E
\left[\Tr_{N_\up N_\down} \hat A  \hat G(E)\right]_c.
\end{align}
The indices $N_\up$ and $N_\down$ indicate the number of $\up, \down$
particles in the connected few-body cluster and therefore $B_{N_\up0}$
and $B_{0N_\down}$ correspond to the ideal-gas contributions from
$\up$ and $\down$ species, respectively.  If both $N_\up$ and
$N_\down$ are finite, $B_{N_\up N_\down}$ is non-zero only when the
interspecies interaction is present.

In the particular case of a spin- and mass-balanced Fermi gas,
i.e., $m\equiv m_\sigma$, $\mu\equiv\mu_\sigma$, $z\equiv z_\sigma$
with $\sigma=\up,\down$, Eq.~(\ref{eq:B_definition}) reduces to
\begin{align}
-\beta\O = \frac{2V}{\lambda^d}\sum_{N\ge1}b_{N}z^N,
\label{eq:grand}
\end{align}
where the thermal wavelength takes its usual form: $\lambda =\lambda_r=
\sqrt{2\pi/mT}$.  The virial coefficients in this case are related to
those occuring in Eq.~\eqref{eq:B_definition} above by $2b_N =
\sum_{N'=0}^{N}B_{N',N-N'}$.  In addition, one may separate out the
effects of interaction, i.e., $b_N = b_N^\text{free}+\Delta b_N$. The
ideal-gas contribution is given by $2b_N^\text{free} = B_{N0}+B_{0N}$
and the effect of interactions are contained in $2\Delta
b_N=\sum_{N'=1}^{N-1}B_{N',N-N'}$.  We note further that in a
mass-balanced system, $B_{NN'}=B_{N'N}$.

It is straightforward to connect the virial coefficients for the
uniform system in Eq.~\eqref{eq:grand} to those for a gas confined in
an isotropic harmonic trap $V(\mathbf{r}) = \frac{1}{2} m \omega^2
r^2$, where $\omega$ is the trapping frequency.  Assuming we are in
the limit where $\beta \omega \ll 1$, the grand potential in the trap
is given by $-\beta\Omega_{\rm trap} = 2 (\beta\omega)^{-d}
\sum_{N\geq 1} b_N^{\rm trap} z^N$.  Applying the local density
approximation to Eq.~\eqref{eq:grand} and comparing expressions then
simply yields $b_N = N^{d/2} \ b_N^{\rm trap}$. This procedure may be
easily extended to the case where $m_\sigma$, $\mu_\sigma$ and/or the
trapping frequencies for each spin are all different.



\subsection{$\mathbf{N_\down=0}$\label{sect:Bfree}}

In this section, we consider the ideal Fermi gas contribution to the
grand potential.  For $N_\up=1$ and $N_\down=0$ the Green's operator
is simply
\begin{align}
\hat G(E)= 
\begin{tikzpicture}[baseline={([yshift=-.6ex]current  bounding  box.center)}]
\draw[blue] (0,0) -- (0.6,0);
\end{tikzpicture},
\end{align}
where the straight line denotes the one-particle propagator. Taking
the trace then requires the state on the left and right to be the
same, i.e.,
\begin{align}
\Tr_{10}\hat A\hat G(E)=
\begin{tikzpicture}[baseline={([yshift=-.6ex]current  bounding  box.center)}]
\node[blue] at (-0.1,0.3) {1};
\draw[blue] (0,0.3) -- (0.4,0.3);
\node[blue] at (0.5,0.3) {1};
\end{tikzpicture}
\end{align}
where the repeated index indicates that the end-points have been
contracted. Thus we write down the virial coefficient 
\begin{align}
B_{10}
= \frac{\lambda_r^d}{V} \oint'_E
\sum_\p\left(E-\epsilon_{\p\up}\right)^{-1}
= \left(\frac{m_\up}{2m_r}\right)^\frac{3}{2}.
\end{align}
Here and in the
following the trace is evaluated in momentum states, with the
single-particle energy $\epsilon_{\p\sigma}=p^2/2m_\sigma$.

Next we let $N_\up=2$ and write down the Green's operator 
\begin{align}
\hat G(E)= 
\begin{tikzpicture}[baseline={([yshift=-.6ex]current  bounding  box.center)}]
\draw[blue] (0,0.3) -- (0.6,0.3);
\draw[blue] (0,0) -- (0.6,0);
\end{tikzpicture}.
\end{align}
Since there is no intraspecies interaction, each particle can only
propagate as a free particle. Thus, few- and many-body correlations
can only arise through exchange of identical particles.  Applying the
exchange operator and taking the trace, we obtain
\begin{align}
\Tr_{20}\hat A\hat G(E)=\frac{1}{2!} \left(
\begin{tikzpicture}[baseline={([yshift=-.6ex]current  bounding  box.center)}]
\node[blue] at (-0.1,0.3) {1};
\node[blue] at (-0.1,0) {2};
\draw[blue] (0,0.3) -- (0.4,0.3);
\draw[blue] (0,0) -- (0.4,0);
\node[blue] at (0.5,0.3) {1};
\node[blue] at (0.5,0) {2};
\end{tikzpicture}
-
\begin{tikzpicture}[baseline={([yshift=-.6ex]current  bounding  box.center)}]
\node[blue] at (-0.1,0.3) {1};
\node[blue] at (-0.1,0) {2};
\draw[blue] (0,0.3) -- (0.4,0.3);
\draw[blue] (0,0) -- (0.4,0);
\node[blue] at (0.5,0.3) {2};
\node[blue] at (0.5,0) {1};
\end{tikzpicture}\right).
\end{align}
The factor $1/2!$ accounts for the fact that there are two identical
clusters (one-particle propagators) in the operator $\hat G$.  We see
that the two propagators in the first term are not connected and thus
only the second term contributes to the virial coefficient
\begin{align}
B_{20}
= \frac{\lambda_r^d}{V} \oint'_E
\sum_\p\frac{-1}{2!} \left(E - 2\epsilon_{\p\up}\right)^{-1}
=\left(\frac{m_\up}{2m_r}\right)^\frac{3}{2}\frac{-1}{2\cdot2^{\frac{d}{2}}},
\nn
\end{align}
Note that the consequence of the trace is that each particle carries
the same momentum.

For $N_\up = 3$, there are $3!$ permutations,
two of which are connected; hence,
\begin{align}
[\Tr_{30}\hat A\hat G(E)]_c=
\frac{1}{3!} \left(
\begin{tikzpicture}[baseline={([yshift=-.6ex]current  bounding  box.center)}]
\node[blue] at (-0.1,0.6) {1};
\node[blue] at (-0.1,0.3) {2};
\node[blue] at (-0.1,0) {3};
\draw[blue] (0,0.6) -- (0.4,0.6);
\draw[blue] (0,0.3) -- (0.4,0.3);
\draw[blue] (0,0) -- (0.4,0);
\node[blue] at (0.5,0.6) {3};
\node[blue] at (0.5,0.3) {1};
\node[blue] at (0.5,0) {2};
\end{tikzpicture}
+
\begin{tikzpicture}[baseline={([yshift=-.6ex]current  bounding  box.center)}]
\node[blue] at (-0.1,0.6) {1};
\node[blue] at (-0.1,0.3) {2};
\node[blue] at (-0.1,0) {3};
\draw[blue] (0,0.6) -- (0.4,0.6);
\draw[blue] (0,0.3) -- (0.4,0.3);
\draw[blue] (0,0) -- (0.4,0);
\node[blue] at (0.5,0.6) {2};
\node[blue] at (0.5,0.3) {3};
\node[blue] at (0.5,0) {1};
\end{tikzpicture}\right)
=
\frac{2}{3!}\cdot
\begin{tikzpicture}[baseline={([yshift=-.6ex]current  bounding  box.center)}]
\node[blue] at (-0.1,0.6) {1};
\node[blue] at (-0.1,0.3) {2};
\node[blue] at (-0.1,0) {3};
\draw[blue] (0,0.6) -- (0.4,0.6);
\draw[blue] (0,0.3) -- (0.4,0.3);
\draw[blue] (0,0) -- (0.4,0);
\node[blue] at (0.5,0.6) {3};
\node[blue] at (0.5,0.3) {1};
\node[blue] at (0.5,0) {2};
\end{tikzpicture}
\nn
\end{align}
This leads to
\begin{align}
B_{30}
= \frac{\lambda_r^d}{V} \oint'_E
\sum_\p\frac{2}{3!} \left(E - 3\epsilon_{\p\up}\right)^{-1}
=\left(\frac{m_\up}{2m_r}\right)^\frac{3}{2}\frac{1}{3\cdot3^{\frac{d}{2}}}.
\nn
\end{align}

It is easy to verify that for any $N_\up>0$, there are $(N_\up-1)!$
connected diagrams each with identical contribution and permutation
sign $(-1)^{N_\up-1}$.  Hence, the virial coefficients of a free Fermi
gas are given by
\begin{align}
  B_{N_\up 0}=\left(\frac{m_\up}{2m_r}\right)^\frac{3}{2}
  \frac{(-1)^{N_\up-1}}{N_\up\cdot N_\up^{\frac{d}{2}}}.
\label{eq:bfree}
\end{align}

In a spin- and mass-balanced Fermi gas, Eq.~\eqref{eq:bfree} yields
$b_N^{\text{free}}= \frac{1}{2}(B_{N0}+B_{0N}) =
\frac{(-1)^{N-1}}{N\cdot N^{d/2}}$, as expected.



\subsection{$\mathbf{N_\up=N_\down=1}$\label{sect:B11}}

In order to deal with interactions, we use the fact that the full Green's
operator can be written in terms of the free Green's operator and the
$T$ matrix, defined such that $\hat T^{-1}(E) = \hat H_\text{int}^{-1}
- \hat G_0(E)$. For the case of $N_\up=N_\down=1$, we write down the
Green's operator
\begin{align}
\hat G(E)= \hat G_0(E) +\hat G_0(E) \hat T_2(E) \hat G_0(E) =
\begin{tikzpicture}[baseline={([yshift=-.6ex]current  bounding  box.center)}]
\draw[blue] (0,0.3) -- (0.4,0.3);
\draw[red] (0,0) -- (0.4,0);
\end{tikzpicture}
+
\begin{tikzpicture}[baseline={([yshift=-.6ex]current  bounding  box.center)}]
\draw[T2] (0.2,0.15) -- (0.4,0.15);
\draw[blue] (0,0.3) -- (0.2,0.15);
\draw[red] (0.2,0.15)--(0,0);
\draw[blue]  (0.4,0.15) -- (0.6,0.3);
\draw[red]  (0.6,0) -- (0.4,0.15);
\end{tikzpicture}
,
\end{align}
where $\tikz[baseline={([yshift=-.57ex]current bounding box.center)}]
\draw[T2] (0,0) -- (0.4,0); $ denotes the two-body $T$ operator $\hat
T_2(E)$ and particles of different species are depicted by different
colours. Then, the trace of $\hat G$ is given by
\begin{align}
\Tr_{11}\hat G(E) 
=
\begin{tikzpicture}[baseline={([yshift=-.6ex]current  bounding  box.center)}]
\node[blue] at (-0.1,0.3) {1};
\node[red] at (-0.1,0) {1};
\draw[blue] (0,0.3) -- (0.4,0.3);
\draw[red] (0,0) -- (0.4,0);
\node[blue] at (0.5,0.3) {1};
\node[red] at (0.5,0) {1};
\end{tikzpicture}
+
\begin{tikzpicture}[baseline={([yshift=-.6ex]current  bounding  box.center)}]
\node[blue] at (-0.1,0.3) {1};
\node[red] at (-0.1,0) {1};
\draw[T2] (0.2,0.15) -- (0.4,0.15);
\draw[blue] (0,0.3) -- (0.2,0.15);
\draw[red] (0.2,0.15)--(0,0);
\draw[blue]  (0.4,0.15) -- (0.6,0.3);
\draw[red]  (0.6,0) -- (0.4,0.15);
\node[blue] at (0.7,0.3) {1};
\node[red] at (0.7,0) {1};
\end{tikzpicture}
.
\end{align}
Since the two particles are distinguishable there are no additional
exchange terms. 
Discarding the disconnected diagram, one obtains
\begin{align}
[\Tr_{11}\hat G(E)]_c =
\begin{tikzpicture}[baseline={([yshift=-.6ex]current  bounding  box.center)}]
\node[blue] at (-0.1,0.3) {1};
\node[red] at (-0.1,0) {1};
\draw[T2] (0.2,0.15) -- (0.4,0.15);
\draw[blue] (0,0.3) -- (0.2,0.15);
\draw[red] (0.2,0.15)--(0,0);
\draw[blue]  (0.4,0.15) -- (0.6,0.3);
\draw[red]  (0.6,0) -- (0.4,0.15);
\node[blue] at (0.7,0.3) {1};
\node[red] at (0.7,0) {1};
\end{tikzpicture}
&= \Tr_{11}  \hat G_0(E)\hat T_2(E)\hat G_0(E) 
\\&= \Tr_{11}  \hat T_2(E)\frac{d}{dE}\hat T^{-1}_2(E) \label{connectedTr11},
\end{align}
where in the last step we used the cyclic property of the trace and
the identity: $\frac{d}{dE}\hat T^{-1}(E) = -\frac{d}{dE}\hat G_0(E) =
\hat G_0^2(E)$.

The two-body $T$ matrix is given by
\begin{align}
\hat T_2(E) =  \sum_{\P} \sum_{\ell,\ell_z} 
T_{2,\ell}(E-P^2/2M)
\ket{\P,\ell}\bra{\P,\ell},
\end{align}
where $\P$ denotes the centre-of-mass momentum, $\ell$ ($\ell_z$) the
total (magnetic) angular momentum quantum number and
$M=m_\up+m_\down$ is the total mass. The $T$ matrix is independent of
the magnetic quantum number, resulting in the degeneracy factor
$\xi_\ell \equiv \sum_{\ell_z}$: This takes the value $2\ell+1$ in 3D,
$2-\delta_{0,\ell}$ in 2D, and 1 in 1D (where $\ell$ is restricted to
0 or 1).
Both $\P$ and $\ell$ are conserved quantities and thus $\hat T_2$ is
diagonal in this representation. The inverse of this operator is thus
given by
\begin{align}
\hat T^{-1}_2(E) = \sum_{\P,\ell} \xi_\ell 
T^{-1}_{2,\ell}(E-P^2/2M)
\ket{\P,\ell}\bra{\P,\ell}.
\end{align}
As a result, Eq.~(\ref{connectedTr11}) reads
\begin{align}
\begin{tikzpicture}[baseline={([yshift=-.6ex]current  bounding  box.center)}]
\node[blue] at (-0.1,0.3) {1};
\node[red] at (-0.1,0) {1};
\draw[T2] (0.2,0.15) -- (0.4,0.15);
\draw[blue] (0,0.3) -- (0.2,0.15);
\draw[red] (0.2,0.15)--(0,0);
\draw[blue]  (0.4,0.15) -- (0.6,0.3);
\draw[red]  (0.6,0) -- (0.4,0.15);
\node[blue] at (0.7,0.3) {1};
\node[red] at (0.7,0) {1};
\end{tikzpicture}
= \sum_{\P,\ell}\xi_\ell  T_{2,\ell}\left(E-\frac{P^2}{2M}\right)\frac{d}{dE} T^{-1}_{2,\ell}\left(E-\frac{P^2}{2M}\right) .
\end{align}
Hence, the second virial coefficient is given by
\begin{align}
  B_{11}= \left(\frac{M}{2m_r}\right)^{\frac{d}{2}}\oint'_E\sum_\ell
  \xi_\ell T_{2,\ell}(E)\frac{d}{d E}T^{-1}_{2,\ell}(E)\label{eq:B11},
\end{align}
which is identical to that obtained in Ref.~\cite{Kaplan2011}. The
pre-factor results from integrating out the centre-of-mass motion,
i.e., $(\lambda_r^3/V)\sum_\P \exp[-\beta\frac{P^2}{2M}] =
(\frac{M}{2m_r})^{3/2}$.

The second virial coefficient \eqref{eq:B11} can be shown to be
equivalent to the well-known Beth-Uhlenbeck formula. To see this, we
consider the contributions from the bound states and the scattering
states separately.  The former is easy to extract.  Near an energy
pole $E=-\eb$, the $T$ matrix is proportional to $(E+\eb)^{-1}$ and
thus $T_{2,\ell}(E)\frac{d}{dE}T^{-1}_{2,\ell}(E) \rightarrow
(E+\eb)^{-1}$.  As a result, the contour integral picks up this simple
pole to give
\begin{align}
B_{11}^\text{poles}=
\left(\frac{M}{2m_r}\right)^{\frac{d}{2}}\sum_{\text{b}}e^{\beta\varepsilon_{\text{b}}},
\nn
\end{align}
where the sum is over all bound states -- even if these are degenerate,
they should be counted separately. To tackle the scattering states, we
deform the contour to the real axis, yielding (mass ratio factor and
partial-wave sum omitted)
\begin{align}
-\int_0^\infty \frac{dE}{\pi} e^{-\beta E} 
\Im \left[\frac{d}{d E}\ln T^{-1}_{2,\ell}(E+i0)\right].
\nn
\end{align}
Noting that $T_{2,\ell}(E+i0)\propto(k\cot \delta_\ell(k)-ik)^{-1}$ with 
$k=\sqrt{2m_rE}$, the integral above reduces to 
\begin{align}
\int_0^\infty \frac{dE}{\pi} e^{-\beta E} 
\frac{dk}{dE} \delta'_\ell(k).
\nn
\end{align}
Replacing $E$ with $k^2/2m_r$, we recover the Beth-Uhlenbeck
formula (see, e.g., \cite{Pathria})
\begin{align}\label{eq:B-Uh}
B_{11} = \left(\frac{M}{2m_r}\right)^{\frac{d}{2}}\left[
\sum_{\text{b}}e^{\beta\varepsilon_{\text{b}}} + 
\sum_\ell\xi_\ell \int_0^\infty \frac{dk}{\pi} 
e^{-\frac{\beta k^2}{2m_r}} \delta'_\ell(k)
\right] \! .
\end{align}

Finally, we note that in the spin- and mass-balanced system, the part
of the virial coefficient arising from interactions is obtained from
the relation $2\Delta b_2 = B_{11}$.



\subsection{$\mathbf{N_\up=2; \,N_\down=1}$\label{sect:B21}}

In this section, we consider the contribution to the grand potential
from the few-body cluster with two $\up$ and one $\down$ atoms.
In this case the Green's operator reads
\begin{align}
\hat G(E) = 
\begin{tikzpicture}[baseline={([yshift=-.6ex]current  bounding  box.center)}]
\draw[blue] (0,-0.1) -- (0.4,-0.1);\draw[red] (0,0.1) -- (0.4,0.1) ;\draw[blue] (0,0.3) -- (0.4,0.3);
\end{tikzpicture}
+
\begin{tikzpicture}[baseline={([yshift=-.6ex]current  bounding  box.center)}]
\draw[red] (0,0.1) -- (0.2,-0.1);\draw[blue] (0.2,-0.1) -- (0,-0.1);
\draw[red] (0.6,0.1) -- (0.4,-0.1);\draw[blue] (0.4,-0.1) -- (0.6,-0.1);
\draw[T2] (0.2,-0.1) -- (0.4,-0.1); \draw[blue] (0,0.3) -- (0.6,0.3);
\end{tikzpicture}
+
\begin{tikzpicture}[baseline={([yshift=-.6ex]current  bounding  box.center)}]
\draw[red] (0,0.1) -- (0.2,-0.1);\draw[blue] (0.2,-0.1) -- (0,-0.1);
\draw[T2] (0.2,-0.1) -- (0.4,-0.1); \draw[blue] (0,0.3) -- (0.6,0.3);
\draw[red] (0.6,0.3) -- (0.4,-0.1);\draw[blue] (0.4,-0.1) -- (1,-0.1);
\draw[T2] (0.6,0.3) -- (0.8,0.3);
\draw[blue] (1,0.3) -- (0.8,0.3);\draw[red] (0.8,0.3) -- (1,0.1);
\end{tikzpicture}
+
\begin{tikzpicture}[baseline={([yshift=-.6ex]current  bounding  box.center)}]
\draw[red] (0,0.1) -- (0.2,-0.1);\draw[blue] (0.2,-0.1) -- (0,-0.1);
\draw[T2] (0.2,-0.1) -- (0.4,-0.1); \draw[blue] (0,0.3) -- (0.6,0.3);
\draw[red] (0.6,0.3) -- (0.4,-0.1);\draw[blue] (0.4,-0.1) -- (1,-0.1);
\draw[T2] (0.6,0.3) -- (0.8,0.3);
\draw[blue] (1.4,0.3) -- (0.8,0.3);
\draw[red] (0.8,0.3) -- (1,-0.1);
\draw[T2] (1,-0.1) -- (1.2,-0.1);
\draw[blue] (1.4,-0.1) -- (1.2,-0.1);\draw[red] (1.2,-0.1) -- (1.4,0.1);
\end{tikzpicture}
+\dots
\end{align}
We see that the first diagram corresponds to free propagation and is
disconnected. The second term appears, at first glance, to be
disconnected but we shall see that the one-particle and two-particle
sub-clusters can be connected via exchange. Since the largest
sub-cluster in the second term contains two particles, we will call
this term the `two-body' contribution. The remaining terms are
fully connected and thus we refer to these collectively as the
`three-body' contributions.

In the next step, we apply the permutation operator and take the trace.
After discarding the free propagator, we have
\begin{align}
[\Tr_{21}\hat A\hat G(E)]_\text{int} = \hspace{5.cm}&
\nn\\
\left(
\begin{tikzpicture}[baseline={([yshift=-.6ex]current  bounding  box.center)}]
\draw[red] (0,0.1) -- (0.2,-0.1);\draw[blue] (0.2,-0.1) -- (0,-0.1);
\draw[red] (0.6,0.1) -- (0.4,-0.1);\draw[blue] (0.4,-0.1) -- (0.6,-0.1);
\draw[T2] (0.2,-0.1) -- (0.4,-0.1); \draw[blue] (0,0.3) -- (0.6,0.3);
\node[blue] at (-0.1,-0.1) {1};\node at (-0.1,0.1) [red]{1};\node[blue] at (-0.1,0.3) {2};
\node[blue] at (0.7,-0.1) {1};\node at (0.7,0.1) [red]{1};\node[blue] at (0.7,0.3) {2};
\end{tikzpicture}
+
\begin{tikzpicture}[baseline={([yshift=-.6ex]current  bounding  box.center)}]
\draw[red] (0,0.1) -- (0.2,-0.1);\draw[blue] (0.2,-0.1) -- (0,-0.1);
\draw[T2] (0.2,-0.1) -- (0.4,-0.1); \draw[blue] (0,0.3) -- (0.6,0.3);
\draw[red] (0.6,0.3) -- (0.4,-0.1);\draw[blue] (0.4,-0.1) -- (1,-0.1);
\draw[T2] (0.6,0.3) -- (0.8,0.3);
\draw[blue] (1,0.3) -- (0.8,0.3);\draw[red] (0.8,0.3) -- (1,0.1);
\node[blue] at (-0.1,-0.1) {1};\node at (-0.1,0.1) [red]{1};\node[blue] at (-0.1,0.3) {2};
\node[blue] at (1.1,-0.1) {1};\node at (1.1,0.1) [red]{1};\node[blue] at (1.1,0.3) {2};
\end{tikzpicture}
+
\begin{tikzpicture}[baseline={([yshift=-.6ex]current  bounding  box.center)}]
\node[blue] at (-0.1,-0.1) {1};\node at (-0.1,0.1) [red]{1};\node[blue] at (-0.1,0.3) {2};
\draw[red] (0,0.1) -- (0.2,-0.1);\draw[blue] (0.2,-0.1) -- (0,-0.1);
\draw[T2] (0.2,-0.1) -- (0.4,-0.1); \draw[blue] (0,0.3) -- (0.6,0.3);
\draw[red] (0.6,0.3) -- (0.4,-0.1);\draw[blue] (0.4,-0.1) -- (1,-0.1);
\draw[T2] (0.6,0.3) -- (0.8,0.3);
\draw[blue] (1.4,0.3) -- (0.8,0.3);
\draw[red] (0.8,0.3) -- (1,-0.1);
\draw[T2] (1,-0.1) -- (1.2,-0.1);
\draw[blue] (1.4,-0.1) -- (1.2,-0.1);\draw[red] (1.2,-0.1) -- (1.4,0.1);
\node[blue] at (1.5,-0.1) {1};\node at (1.5,0.1) [red]{1};\node[blue] at (1.5,0.3) {2};
\end{tikzpicture}
+\dots\right)&
\nn\\
-\left(
\begin{tikzpicture}[baseline={([yshift=-.6ex]current  bounding  box.center)}]
\draw[red] (0,0.1) -- (0.2,-0.1);\draw[blue] (0.2,-0.1) -- (0,-0.1);
\draw[red] (0.6,0.1) -- (0.4,-0.1);\draw[blue] (0.4,-0.1) -- (0.6,-0.1);
\draw[T2] (0.2,-0.1) -- (0.4,-0.1); \draw[blue] (0,0.3) -- (0.6,0.3);
\node[blue] at (-0.1,-0.1) {1};\node at (-0.1,0.1) [red]{1};\node[blue] at (-0.1,0.3) {2};
\node[blue] at (0.7,-0.1) {2};\node at (0.7,0.1) [red]{1};\node[blue] at (0.7,0.3) {1};
\end{tikzpicture}
+
\begin{tikzpicture}[baseline={([yshift=-.6ex]current  bounding  box.center)}]
\draw[red] (0,0.1) -- (0.2,-0.1);\draw[blue] (0.2,-0.1) -- (0,-0.1);
\draw[T2] (0.2,-0.1) -- (0.4,-0.1); \draw[blue] (0,0.3) -- (0.6,0.3);
\draw[red] (0.6,0.3) -- (0.4,-0.1);\draw[blue] (0.4,-0.1) -- (1,-0.1);
\draw[T2] (0.6,0.3) -- (0.8,0.3);
\draw[blue] (1,0.3) -- (0.8,0.3);\draw[red] (0.8,0.3) -- (1,0.1);
\node[blue] at (-0.1,-0.1) {1};\node at (-0.1,0.1) [red]{1};\node[blue] at (-0.1,0.3) {2};
\node[blue] at (1.1,-0.1) {2};\node at (1.1,0.1) [red]{1};\node[blue] at (1.1,0.3) {1};
\end{tikzpicture}
+
\begin{tikzpicture}[baseline={([yshift=-.6ex]current  bounding  box.center)}]
\node[blue] at (-0.1,-0.1) {1};\node at (-0.1,0.1) [red]{1};\node[blue] at (-0.1,0.3) {2};
\draw[red] (0,0.1) -- (0.2,-0.1);\draw[blue] (0.2,-0.1) -- (0,-0.1);
\draw[T2] (0.2,-0.1) -- (0.4,-0.1); \draw[blue] (0,0.3) -- (0.6,0.3);
\draw[red] (0.6,0.3) -- (0.4,-0.1);\draw[blue] (0.4,-0.1) -- (1,-0.1);
\draw[T2] (0.6,0.3) -- (0.8,0.3);
\draw[blue] (1.4,0.3) -- (0.8,0.3);
\draw[red] (0.8,0.3) -- (1,-0.1);
\draw[T2] (1,-0.1) -- (1.2,-0.1);
\draw[blue] (1.4,-0.1) -- (1.2,-0.1);\draw[red] (1.2,-0.1) -- (1.4,0.1);
\node[blue] at (1.5,-0.1) {2};\node at (1.5,0.1) [red]{1};\node[blue] at (1.5,0.3) {1};
\end{tikzpicture}
+\dots
\right)&. \nn
\end{align}
The three-body parts of the series in the above equation may be
replaced by three-body propagators, but first it is instructive to
reorganize the diagrams into two groups: (i) diagrams where the
$T_2$'s on the left and right are labelled by the same indices and
(ii) those where they are labelled by different indices. Following the
reorganization, the two sets of diagrams are
\begin{align}
[\Tr_{21}\hat A\hat G(E)]_\text{int} =\hspace{5cm}&
\nn\\ \nn 
\left(
\begin{tikzpicture}[baseline={([yshift=-.6ex]current  bounding  box.center)}]
\draw[red] (0,0.1) -- (0.2,-0.1);\draw[blue] (0.2,-0.1) -- (0,-0.1);
\draw[red] (0.6,0.1) -- (0.4,-0.1);\draw[blue] (0.4,-0.1) -- (0.6,-0.1);
\draw[T2] (0.2,-0.1) -- (0.4,-0.1); \draw[blue] (0,0.3) -- (0.6,0.3);
\node[blue] at (-0.1,-0.1) {1};\node at (-0.1,0.1) [red]{1};\node[blue] at (-0.1,0.3) {2};
\node[blue] at (0.7,-0.1) {1};\node at (0.7,0.1) [red]{1};\node[blue] at (0.7,0.3) {2};
\end{tikzpicture}
-
\begin{tikzpicture}[baseline={([yshift=-.6ex]current  bounding  box.center)}]
\draw[red] (0,0.1) -- (0.2,-0.1);\draw[blue] (0.2,-0.1) -- (0,-0.1);
\draw[T2] (0.2,-0.1) -- (0.4,-0.1); \draw[blue] (0,0.3) -- (0.6,0.3);
\draw[red] (0.6,0.3) -- (0.4,-0.1);\draw[blue] (0.4,-0.1) -- (1,-0.1);
\draw[T2] (0.6,0.3) -- (0.8,0.3);
\draw[blue] (1,0.3) -- (0.8,0.3);\draw[red] (0.8,0.3) -- (1,0.1);
\node[blue] at (-0.1,-0.1) {1};\node at (-0.1,0.1) [red]{1};\node[blue] at (-0.1,0.3) {2};
\node[blue] at (1.1,-0.1) {2};\node at (1.1,0.1) [red]{1};\node[blue] at (1.1,0.3) {1};
\end{tikzpicture}
+
\begin{tikzpicture}[baseline={([yshift=-.6ex]current  bounding  box.center)}]
\node[blue] at (-0.1,-0.1) {1};\node at (-0.1,0.1) [red]{1};\node[blue] at (-0.1,0.3) {2};
\draw[red] (0,0.1) -- (0.2,-0.1);\draw[blue] (0.2,-0.1) -- (0,-0.1);
\draw[T2] (0.2,-0.1) -- (0.4,-0.1); \draw[blue] (0,0.3) -- (0.6,0.3);
\draw[red] (0.6,0.3) -- (0.4,-0.1);\draw[blue] (0.4,-0.1) -- (1,-0.1);
\draw[T2] (0.6,0.3) -- (0.8,0.3);
\draw[blue] (1.4,0.3) -- (0.8,0.3);
\draw[red] (0.8,0.3) -- (1,-0.1);
\draw[T2] (1,-0.1) -- (1.2,-0.1);
\draw[blue] (1.4,-0.1) -- (1.2,-0.1);\draw[red] (1.2,-0.1) -- (1.4,0.1);
\node[blue] at (1.5,-0.1) {1};\node at (1.5,0.1) [red]{1};\node[blue] at (1.5,0.3) {2};
\end{tikzpicture}
-\dots\right)& 
\\
-\left(
\begin{tikzpicture}[baseline={([yshift=-.6ex]current  bounding  box.center)}]
\draw[red] (0,0.1) -- (0.2,-0.1);\draw[blue] (0.2,-0.1) -- (0,-0.1);
\draw[red] (0.6,0.1) -- (0.4,-0.1);\draw[blue] (0.4,-0.1) -- (0.6,-0.1);
\draw[T2] (0.2,-0.1) -- (0.4,-0.1); \draw[blue] (0,0.3) -- (0.6,0.3);
\node[blue] at (-0.1,-0.1) {1};\node at (-0.1,0.1) [red]{1};\node[blue] at (-0.1,0.3) {2};
\node[blue] at (0.7,-0.1) {2};\node at (0.7,0.1) [red]{1};\node[blue] at (0.7,0.3) {1};
\end{tikzpicture}
-
\begin{tikzpicture}[baseline={([yshift=-.6ex]current  bounding  box.center)}]
\draw[red] (0,0.1) -- (0.2,-0.1);\draw[blue] (0.2,-0.1) -- (0,-0.1);
\draw[T2] (0.2,-0.1) -- (0.4,-0.1); \draw[blue] (0,0.3) -- (0.6,0.3);
\draw[red] (0.6,0.3) -- (0.4,-0.1);\draw[blue] (0.4,-0.1) -- (1,-0.1);
\draw[T2] (0.6,0.3) -- (0.8,0.3);
\draw[blue] (1,0.3) -- (0.8,0.3);\draw[red] (0.8,0.3) -- (1,0.1);
\node[blue] at (-0.1,-0.1) {1};\node at (-0.1,0.1) [red]{1};\node[blue] at (-0.1,0.3) {2};
\node[blue] at (1.1,-0.1) {1};\node at (1.1,0.1) [red]{1};\node[blue] at (1.1,0.3) {2};
\end{tikzpicture}
+
\begin{tikzpicture}[baseline={([yshift=-.6ex]current  bounding  box.center)}]
\node[blue] at (-0.1,-0.1) {1};\node at (-0.1,0.1) [red]{1};\node[blue] at (-0.1,0.3) {2};
\draw[red] (0,0.1) -- (0.2,-0.1);\draw[blue] (0.2,-0.1) -- (0,-0.1);
\draw[T2] (0.2,-0.1) -- (0.4,-0.1); \draw[blue] (0,0.3) -- (0.6,0.3);
\draw[red] (0.6,0.3) -- (0.4,-0.1);\draw[blue] (0.4,-0.1) -- (1,-0.1);
\draw[T2] (0.6,0.3) -- (0.8,0.3);
\draw[blue] (1.4,0.3) -- (0.8,0.3);
\draw[red] (0.8,0.3) -- (1,-0.1);
\draw[T2] (1,-0.1) -- (1.2,-0.1);
\draw[blue] (1.4,-0.1) -- (1.2,-0.1);\draw[red] (1.2,-0.1) -- (1.4,0.1);
\node[blue] at (1.5,-0.1) {2};\node at (1.5,0.1) [red]{1};\node[blue] at (1.5,0.3) {1};
\end{tikzpicture}
-\dots
\right)&. \nn
\end{align}
It is clear that the left-most term in the first set is 
disconnected and therefore does not contribute to the virial
coefficient.

We then define the three-body $T$ matrix $t_3$:
\begin{align}
\begin{tikzpicture}[baseline={([yshift=-.6ex]current  bounding  box.center)}]
\draw[blue] (0.2,-0.1) -- (0,-0.1);
\draw[red] (0,0.1) -- (0.2,-0.1);
\draw[T2] (0.2,-0.1) -- (0.4,-0.1); \draw[blue] (0,0.3) -- (0.4,0.3);
\fill[gray,opacity=0.5] (0.4,-0.1) rectangle (0.8,0.3) ;
\node at (0.6,0.1) {$t_3$};
\draw[blue] (0.8,-0.1) -- (1.2,-0.1) ;
\draw[T2] (0.8,0.3) -- (1,0.3);
\draw[blue] (1.2,0.3) -- (1,0.3);\draw[red] (1,0.3) -- (1.2,0.1);
\end{tikzpicture}
\equiv
- 
\begin{tikzpicture}[baseline={([yshift=-.6ex]current  bounding  box.center)}]
\draw[red] (0,0.1) -- (0.2,-0.1);\draw[blue] (0.2,-0.1) -- (0,-0.1);
\draw[T2] (0.2,-0.1) -- (0.4,-0.1); \draw[blue] (0,0.3) -- (0.6,0.3);
\draw[red] (0.6,0.3) -- (0.4,-0.1);\draw[blue] (0.4,-0.1) -- (1,-0.1);
\draw[T2] (0.6,0.3) -- (0.8,0.3);
\draw[blue] (1,0.3) -- (0.8,0.3);\draw[red] (0.8,0.3) -- (1,0.1);
\end{tikzpicture}
+
\begin{tikzpicture}[baseline={([yshift=-.6ex]current  bounding  box.center)}]
\draw[red] (0,0.1) -- (0.2,-0.1);\draw[blue] (0.2,-0.1) -- (0,-0.1);
\draw[T2] (0.2,-0.1) -- (0.4,-0.1); \draw[blue] (0,0.3) -- (0.6,0.3);
\draw[red] (0.6,0.3) -- (0.4,-0.1);\draw[blue] (0.4,-0.1) -- (1,-0.1);
\draw[T2] (0.6,0.3) -- (0.8,0.3);
\draw[blue] (1.4,0.3) -- (0.8,0.3);
\draw[red] (0.8,0.3) -- (1,-0.1);
\draw[T2] (1,-0.1) -- (1.2,-0.1);
\draw[blue] (1.4,-0.1) -- (1.2,-0.1);\draw[red] (1.2,-0.1) -- (1.4,0.1);
\end{tikzpicture}
-\dots,
\label{3bodyTmatrix}
\end{align}
where $t_3$ can be obtained from solving the
Skorniakov--Ter-Martirosian integral equation \cite{stm}, properly
generalized to the present problem. For details, see Appendix
\ref{app:STM}. Hence, the virial coefficient reads
\begin{align}
B_{21} &= \frac{\lambda_r^d}{V}\oint'_E
\left(
\begin{tikzpicture}[baseline={([yshift=-.6ex]current  bounding  box.center)}]
\draw[blue] (0.2,-0.1) -- (0,-0.1);
\draw[red] (0,0.1) -- (0.2,-0.1);
\draw[T2] (0.2,-0.1) -- (0.4,-0.1); \draw[blue] (0,0.3) -- (0.4,0.3);
\fill[gray,opacity=0.5] (0.4,-0.1) rectangle (0.8,0.3) ;
\node at (0.6,0.1) {$t_3$};
\draw[blue] (0.8,-0.1) -- (1.2,-0.1) ;
\draw[T2] (0.8,0.3) -- (1,0.3);
\draw[blue] (1.2,0.3) -- (1,0.3);\draw[red] (1,0.3) -- (1.2,0.1);
\node[blue] at (-0.1,-0.1) {1};\node at (-0.1,0.1) [red]{1};\node[blue] at (-0.1,0.3) {2};
\node[blue] at (1.3,-0.1) {2};\node at (1.3,0.1) [red]{1};\node[blue] at (1.3,0.3) {1};
\end{tikzpicture}
-
\begin{tikzpicture}[baseline={([yshift=-.6ex]current  bounding  box.center)}]
\draw[red] (0,0.1) -- (0.2,-0.1);\draw[blue] (0.2,-0.1) -- (0,-0.1);
\draw[red] (0.6,0.1) -- (0.4,-0.1);\draw[blue] (0.4,-0.1) -- (0.6,-0.1);
\draw[T2] (0.2,-0.1) -- (0.4,-0.1); \draw[blue] (0,0.3) -- (0.6,0.3);
\node[blue] at (-0.1,-0.1) {1};\node at (-0.1,0.1) [red]{1};\node[blue] at (-0.1,0.3) {2};
\node[blue] at (0.7,-0.1) {2};\node at (0.7,0.1) [red]{1};\node[blue] at (0.7,0.3) {1};
\end{tikzpicture}
-
\begin{tikzpicture}[baseline={([yshift=-.6ex]current  bounding  box.center)}]
\draw[blue] (0.2,-0.1) -- (0,-0.1);
\draw[red] (0,0.1) -- (0.2,-0.1);
\draw[T2] (0.2,-0.1) -- (0.4,-0.1); \draw[blue] (0,0.3) -- (0.4,0.3);
\fill[gray,opacity=0.5] (0.4,-0.1) rectangle (0.8,0.3) ;
\node at (0.6,0.1) {$t_3$};
\draw[blue] (0.8,-0.1) -- (1.2,-0.1) ;
\draw[T2] (0.8,0.3) -- (1,0.3);
\draw[blue] (1.2,0.3) -- (1,0.3);\draw[red] (1,0.3) -- (1.2,0.1);
\node[blue] at (-0.1,-0.1) {1};\node at (-0.1,0.1) [red]{1};\node[blue] at (-0.1,0.3) {2};
\node[blue] at (1.3,-0.1) {1};\node at (1.3,0.1) [red]{1};\node[blue] at (1.3,0.3) {2};
\end{tikzpicture}
\right)
\nn\\
&=
\left(\frac{M_{21}}{2m_r}\right)^{\frac{d}{2}}\oint'_E
\sum_{\p}
\Bigg[
 \chi_{21}(\p,\p;E)\frac{d
T^{-1}_2
\left(E-\frac{p^2}{2m_{21}}
\right)}{dE}
\nn\\&\hspace{15mm}
- \frac{T_2(E-p^2/2m_{21})} 
{\left(E-
2\epsilon_{\p\up}-\epsilon_{2\p\down}
\right)^2}
\nn\\&\hspace{15mm}
-\sum_{\p'}
\frac{\chi_{21}(\p,\p';E)}
{\left(E-
\epsilon_{\p\up}-\epsilon_{\p'\up}-\epsilon_{\p+\p'\down}
\right)^2}
\Bigg],
\label{eq:B21}
\end{align}
with $\chi_{21}(\p,\p';E)\equiv
T_2(E-\frac{p^2}{2m_{21}})t_3^{\up\up\down}(\p,\p';E)T_2(E-\frac{p'^2}{2m_{21}})$.
In the last step, the centre-of-mass motion is integrated out, giving
the factor $(\frac{M_{21}}{2m_r})^{\frac{d}{2}}$.  Here, the total
mass is $M_{21} = 2m_\up+m_\down$ and the atom-pair reduced mass
$m_{21}^{-1}=(m_\up+m_\down)^{-1}+m_\up^{-1}$.

Note that the interacting part of the virial coefficient in the spin-
and mass-balanced case is given by $\Delta b_3 =
\frac{1}{2}(B_{21}+B_{12}) = B_{21}$.



\subsection{$\mathbf{N_\up=N_\down=2}$}
 \label{sec:b22}

For two $\up$ and two $\down$ atoms,
we write down the Green's operator,
similarly to the previous cases:
\begin{subequations}
\begin{align}
\hat G(E)
=\,& 
\begin{tikzpicture}[baseline={([yshift=-.55ex]current  bounding  box.center)}]
\draw[blue] (0,-0.1) -- (0.6,-0.1); \draw[blue] (0,0.1) -- (0.6,0.1); 
\draw[red] (0,0.3) -- (0.6,0.3); \draw[red] (0,0.5) -- (0.6,0.5); 
\end{tikzpicture}
+
\begin{tikzpicture}[baseline={([yshift=-.55ex]current  bounding  box.center)}]
\draw[red] (0,0.1) -- (0.2,-0.1) ;\draw[blue](0.2,-0.1) -- (0,-0.1);
\draw[red] (0.6,0.1) -- (0.4,-0.1);\draw[blue] (0.4,-0.1) -- (0.6,-0.1);
\draw[T2] (0.2,-0.1) -- (0.4,-0.1); 
\draw[blue] (0,0.3) -- (0.6,0.3); \draw[red] (0,0.5) -- (0.6,0.5); 
\end{tikzpicture}
+
\begin{tikzpicture}[baseline={([yshift=-.55ex]current  bounding  box.center)}]
\draw[red] (0,0.1) -- (0.2,-0.1);\draw[blue](0.2,-0.1) -- (0,-0.1);
\draw[red] (0.6,0.1) -- (0.4,-0.1);\draw[blue] (0.4,-0.1) -- (0.6,-0.1);
\draw[T2] (0.2,-0.1) -- (0.4,-0.1); 
\draw[blue] (0,0.3) -- (0.2,0.5) ; \draw[red] (0.2,0.5) -- (0,0.5); 
\draw[blue] (0.4,0.5) -- (0.6,0.3); \draw[red] (0.6,0.5) -- (0.4,0.5); 
\draw[T2] (0.2,0.5) -- (0.4,0.5); 
\end{tikzpicture}
\label{eq:B22-2body}\\&
+\left(\,
\begin{tikzpicture}[baseline={([yshift=-.55ex]current  bounding  box.center)}]
\draw[blue] (0.2,-0.1) -- (0,-0.1);\draw[red] (0,0.1) -- (0.2,-0.1);
\draw[T2] (0.2,-0.1) -- (0.4,-0.1); \draw[blue] (0,0.3) -- (0.6,0.3);
\draw[red] (0.6,0.3) -- (0.4,-0.1);\draw[blue] (0.4,-0.1) -- (1,-0.1);
\draw[T2] (0.6,0.3) -- (0.8,0.3);
\draw[red](0.8,0.3) -- (1,0.1);\draw[blue] (1,0.3) -- (0.8,0.3);
 \draw[red] (0,0.5) -- (1,0.5); 
\end{tikzpicture}
+
\begin{tikzpicture}[baseline={([yshift=-.55ex]current  bounding  box.center)}]
\draw[blue](0.2,-0.1) -- (0,-0.1);\draw (0,0.1) -- (0.2,-0.1) [red];
\draw[T2] (0.2,-0.1) -- (0.4,-0.1); \draw[blue] (0,0.3) -- (0.6,0.3);
\draw [blue](0.4,-0.1) -- (1,-0.1);\draw (0.6,0.3) -- (0.4,-0.1)[red];
\draw[T2] (0.6,0.3) -- (0.8,0.3);
\draw (1.4,0.3) -- (0.8,0.3)[blue];\draw[red](0.8,0.3) -- (1,-0.1);
\draw[T2] (1,-0.1) -- (1.2,-0.1);
\draw (1.4,-0.1) -- (1.2,-0.1) [blue];\draw [red](1.2,-0.1) -- (1.4,0.1);
 \draw[red] (0,0.5) -- (1.4,0.5); 
\end{tikzpicture}
+\dots +[\up\leftrightarrow\down]
\right)
\label{eq:B22-3body}\\ 
&
+\left[\,
\begin{tikzpicture}[baseline={([yshift=-.55ex]current  bounding  box.center)}]
\draw[red] (0.2,0.2) -- (0,0.3);\draw (0,0.1) -- (0.2,0.2) [blue];
\draw[T2] (0.2,0.2) -- (0.4,0.2); 
\draw[red] (0,0.5) -- (0.3,0.5) 
	to[in=120,out=0] (0.6,0.2);
\draw[blue]  (0.6,0.2)-- (0.4,0.2);
\draw[red]  (0.4,0.2)
	to[in=180,out=60] (0.7,0.5)
	-- (1.4,0.5);
\draw[T2] (0.6,0.2) -- (0.8,0.2);
\draw[blue] (0,-0.1) -- (0.7,-0.1) 
	to[in=-120,out=0] (1,0.2); 
\draw[red] (1,0.2)-- (0.8,0.2);
\draw [blue](0.8,0.2)
	to[in=180,out=-60] (1.1,-0.1)
	-- (1.4,-0.1); 
\draw[T2] (1,0.2) -- (1.2,0.2); 
\draw[red](1.2,0.2) -- (1.4,0.3);\draw (1.4,0.1) -- (1.2,0.2)[blue];
\end{tikzpicture}
\right.
+
\begin{tikzpicture}[baseline={([yshift=-.55ex]current  bounding  box.center)}]
\draw [blue] (0.2,-0.1) -- (0,-0.1);\draw (0,0.1) -- (0.2,-0.1)[red];
\draw[T2] (0.2,-0.1) -- (0.4,-0.1); 
\draw[blue](0.2,0.5) -- (0,0.3);\draw (0,0.5) -- (0.2,0.5)[red];
\draw[T2] (0.2,0.5) -- (0.4,0.5); 
\draw[red] (0.9,0.5) -- (0.4,0.5);
\draw[blue] (0.4,0.5) -- (0.5,0.2);
\draw[red] (0.5,0.2) -- (0.4,-0.1) ;
\draw[blue] (0.4,-0.1) --  (0.9,-0.1);
\draw[T2] (0.5,0.2) -- (0.7,0.2); 
\draw (0.9,0.3) -- (0.7,0.2)[blue] ;\draw [red](0.7,0.2) -- (0.9,0.1) ;
\end{tikzpicture}
\label{eq:B22-4body-first}\\&\hspace{.5cm}
+
\left(
\begin{tikzpicture}[baseline={([yshift=-.55ex]current  bounding  box.center)}]
\draw[red](0.2,0.2) -- (0,0.3);\draw (0,0.1) -- (0.2,0.2) [blue];
\draw[T2] (0.2,0.2) -- (0.4,0.2); 
\draw[red] (0,0.5) -- (0.3,0.5) 
	to[in=120,out=0] (0.6,0.2);
\draw[blue] (0.6,0.2)-- (0.4,0.2);
\draw[red] (0.4,0.2)
	to[in=180,out=60] (0.7,0.5)
	to[in=120,out=0] (1,0.2);
\draw[blue](1,0.2)-- (0.8,0.2);
\draw[red] (0.8,0.2)
	to[in=180,out=60] (1.1,0.5)
	--(1.8,0.5);
\draw[T2] (0.6,0.2) -- (0.8,0.2);
\draw[blue] (0,-0.1) -- (1.1,-0.1) 
	to[in=-120,out=0] (1.4,0.2);
\draw[red] (1.4,0.2)-- (1.2,0.2);
\draw[blue] (1.2,0.2)
	to[in=180,out=-60] (1.5,-0.1)
	-- (1.8,-0.1); 
\draw[T2] (1,0.2) -- (1.2,0.2); 
\draw[T2] (1.4,0.2) -- (1.6,0.2); 
\draw [red](1.6,0.2) -- (1.8,0.3);\draw (1.8,0.1) -- (1.6,0.2)[blue];
\end{tikzpicture}
+[\up\leftrightarrow\down]
\right)
\label{eq:B22-4body-mid}\\&\hspace{.5cm}
+
\begin{tikzpicture}[baseline={([yshift=-.55ex]current  bounding  box.center)}]
\draw [red](0.2,0.2) -- (0,0.3);\draw (0,0.1) -- (0.2,0.2) [blue];
\draw[T2] (0.2,0.2) -- (0.4,0.2); 
\draw[red] (0,0.5) -- (0.6,0.5);
\draw[blue] (0.6,0.5) -- (0.4,0.2);
\draw[red]  (0.4,0.2) -- (0.6,-0.1) ;
\draw[blue] (0.6,-0.1) -- (0,-0.1);
\draw[T2] (0.6,0.5) -- (0.8,0.5); 
\draw[T2] (0.6,-0.1) -- (0.8,-0.1); 
\draw[T2] (1,0.2) -- (1.2,0.2);
\draw[red] (1.4,0.5) --(0.8,0.5);
\draw[blue] (0.8,0.5) -- (1,0.2) ;
\draw[red]  (1,0.2) -- (0.8,-0.1);
\draw[blue] (0.8,-0.1) -- (1.4,-0.1);
\draw [blue](1.2,0.2)--(1.4,0.1);\draw (1.4,0.3)--(1.2,0.2)[red];
\end{tikzpicture}
+
\left.
\begin{tikzpicture}[baseline={([yshift=-.55ex]current  bounding  box.center)}]
\draw (0,0.1) -- (0.2,-0.1) [red];\draw [blue](0.2,-0.1) -- (0,-0.1);
\draw[T2] (0.2,-0.1) -- (0.4,-0.1); 
\draw (0,0.5) -- (0.2,0.5)[red];\draw [blue](0.2,0.5) -- (0,0.3);
\draw[T2] (0.2,0.5) -- (0.4,0.5); 
\draw[red] (0.4,0.5) -- (0.6,0.5) ;
\draw[blue] (0.6,0.5) -- (0.4,-0.1) ;
\draw[red] (0.4,-0.1) -- (0.6,-0.1) ;
\draw[blue] (0.4,0.5) -- (0.6,-0.1);
\draw[T2] (0.6,0.5) -- (0.8,0.5);
\draw (1,0.5) -- (0.8,0.5)[red];\draw [blue](0.8,0.5) -- (1,0.3);
\draw[T2] (0.6,-0.1) -- (0.8,-0.1);
\draw (1,-0.1) -- (0.8,-0.1)[blue];\draw [red](0.8,-0.1) -- (1,0.1);
\end{tikzpicture}
+\dots
\right]\label{eq:B22-4body-last}.
\end{align}
\end{subequations}
Here, Eq.~(\ref{eq:B22-2body}) depicts the free part and the two-body
contributions; Eq.~(\ref{eq:B22-3body}) the three-body contributions;
and Eqs.~(\ref{eq:B22-4body-first}-\ref{eq:B22-4body-last}) the
four-body contributions. In (\ref{eq:B22-3body}), the infinite
diagrammatic series can be replaced by the three-body propagator,
$t_3$, and thus can be calculated in full. On the other hand, we only
write down explicitly the diagrams containing exactly three $T_2$'s in
Eq.~(\ref{eq:B22-4body-first}) and four $T_2$'s in
Eqs.~(\ref{eq:B22-4body-mid}-\ref{eq:B22-4body-last}). The terms with
five or more $T_2$'s are omitted here and in the following. While not
completely general, this may be shown to correspond to a perturbative
approach (the Born approximation) in certain limits, as discussed in Sec.~\ref{sec:FG}, and
explicitly shown in \ref{sec:vcs}.

To obtain the virial coefficient, one follows the framework outlined
in previous sections. All relevant diagrams can be obtained by keeping
only the connected diagrams resulting from the trace of the Green's
operator under the exchange operator.  We note however that the third
term in Eq.~(\ref{eq:B22-2body}) contains identical two-atom sub-clusters
and thus a factor of $1/2!$ is required to avoid over counting.


\subsection{$\mathbf{N_\up=3; \,N_\down=1}$ \label{sec:b31}}

Here we consider the other possible interacting four-body cluster --
three $\up$ and one $\down$ atoms (there is, of course, an equivalent
diagram with three $\down$ and one $\up$ atoms). As before, we write
down the Green's operator of this few-body system
\begin{subequations}
\begin{align}
\hspace{-3mm}\hat G(E)
=\,&
\begin{tikzpicture}[baseline={([yshift=-.55ex]current  bounding  box.center)}]
\draw[red] (0,-0.1) -- (0.6,-0.1); \draw[blue] (0,0.1) -- (0.6,0.1); 
\draw[blue] (0,0.3) -- (0.6,0.3); \draw[blue] (0,0.5) -- (0.6,0.5); 
\end{tikzpicture}
+
\begin{tikzpicture}[baseline={([yshift=-.55ex]current  bounding  box.center)}]
\draw[blue] (0,0.1) -- (0.2,-0.1);\draw[red] (0.2,-0.1) -- (0,-0.1);
\draw[blue] (0.6,0.1) -- (0.4,-0.1);\draw[red] (0.4,-0.1) -- (0.6,-0.1);
\draw[T2] (0.2,-0.1) -- (0.4,-0.1); 
\draw[blue] (0,0.3) -- (0.6,0.3); \draw[blue] (0,0.5) -- (0.6,0.5); 
\end{tikzpicture}
+
\left[\,
\begin{tikzpicture}[baseline={([yshift=-.55ex]current  bounding  box.center)}]
\draw[red] (0,0.1) -- (0.2,-0.1);\draw[blue] (0.2,-0.1) -- (0,-0.1);
\draw[T2] (0.2,-0.1) -- (0.4,-0.1); \draw[blue] (0,0.3) -- (0.6,0.3);
\draw[red] (0.6,0.3) -- (0.4,-0.1);\draw[blue] (0.4,-0.1) -- (1,-0.1);
\draw[T2] (0.6,0.3) -- (0.8,0.3);
\draw[blue] (1,0.3) -- (0.8,0.3);\draw[red] (0.8,0.3) -- (1,0.1);
\draw[blue] (0,0.5) -- (1,0.5); 
\end{tikzpicture}
+
\begin{tikzpicture}[baseline={([yshift=-.55ex]current  bounding  box.center)}]
\draw[red] (0,0.1) -- (0.2,-0.1);\draw[blue] (0.2,-0.1) -- (0,-0.1);
\draw[T2] (0.2,-0.1) -- (0.4,-0.1); \draw[blue] (0,0.3) -- (0.6,0.3);
\draw[red](0.4,-0.1) -- (0.6,0.3);\draw[blue] (0.4,-0.1) -- (1,-0.1);
\draw[T2] (0.6,0.3) -- (0.8,0.3);
\draw[blue] (1.4,0.3) -- (0.8,0.3);\draw[red] (0.8,0.3) -- (1,-0.1);
\draw[T2] (1,-0.1) -- (1.2,-0.1);
\draw[blue] (1.4,-0.1) -- (1.2,-0.1);\draw[red] (1.2,-0.1) -- (1.4,0.1);
 \draw[blue] (0,0.5) -- (1.4,0.5); 
\end{tikzpicture}
+\dots
\right]
\label{eq:B31-one}
\\&
+\left[\,
\begin{tikzpicture}[baseline={([yshift=-.55ex]current  bounding  box.center)}]
\draw[red] (0,0.1) -- (0.2,0.2) ;\draw[blue] (0.2,0.2) -- (0,0.3);
\draw[T2] (0.2,0.2) -- (0.4,0.2); 
\draw[blue] (0,0.5) -- (0.3,0.5) 
	to[in=120,out=0] (0.6,0.2);
\draw[blue] (0.4,0.2)
	to[in=180,out=60] (0.7,0.5)
	-- (1.4,0.5);
\draw[red]  (0.6,0.2)-- (0.4,0.2);
\draw[T2] (0.6,0.2) -- (0.8,0.2);
\draw[blue] (0,-0.1) -- (0.7,-0.1) 
	to[in=-120,out=0] (1,0.2); 
\draw[blue] (0.8,0.2)
	to[in=180,out=-60] (1.1,-0.1)
	-- (1.4,-0.1); 
\draw[red] (1,0.2)-- (0.8,0.2); 
\draw[T2] (1,0.2) -- (1.2,0.2); 
\draw[red] (1.4,0.1) -- (1.2,0.2);\draw[blue] (1.2,0.2) -- (1.4,0.3);
\end{tikzpicture}
+
\begin{tikzpicture}[baseline={([yshift=-.55ex]current  bounding  box.center)}]
\draw[red] (0,0.1) -- (0.2,0.2);\draw[blue] (0.2,0.2) -- (0,0.3);
\draw[T2] (0.2,0.2) -- (0.4,0.2); 
\draw[blue] (0,0.5) -- (0.3,0.5) 
	to[in=120,out=0] (0.6,0.2);
\draw[red] (0.6,0.2) -- (0.4,0.2);
\draw[blue] (0.4,0.2)
	to[in=180,out=60] (0.7,0.5)
	to[in=120,out=0] (1,0.2);
\draw[red] (1,0.2)-- (0.8,0.2);
\draw[blue] (0.8,0.2)
	to[in=180,out=60] (1.1,0.5)
	--(1.8,0.5);
\draw[T2] (0.6,0.2) -- (0.8,0.2);
\draw[blue] (0,-0.1) -- (1.1,-0.1) 
	to[in=-120,out=0] (1.4,0.2); 
\draw[red] (1.4,0.2) -- (1.2,0.2); 
\draw[blue] (1.2,0.2)
	to[in=180,out=-60] (1.5,-0.1)
	-- (1.8,-0.1); 
\draw[T2] (1,0.2) -- (1.2,0.2); 
\draw[T2] (1.4,0.2) -- (1.6,0.2); 
\draw[red] (1.8,0.1) -- (1.6,0.2);\draw[blue] (1.6,0.2) -- (1.8,0.3);
\end{tikzpicture}
+\dots\right].
\label{eq:B31-two}
\end{align}
\end{subequations}
In Eq.~(\ref{eq:B31-one}), the first term denotes the free propagator;
the second the two-body contribution; and the terms in the bracket the
three-body contributions which can be calculated exactly by means of
the STM integral equation. The four-body contribution is written down
in Eq.~(\ref{eq:B31-two}) where the diagrams containing more than four
$T_2$'s are omitted. Again we note that this can be shown to
correspond to a perturbative approach in certain limits.

All relevant diagrams can be obtained by applying the exchange 
operator to the diagrams above and taking the trace. We note that the 
two-body diagram [the second term in Eq.~(\ref{eq:B31-one})] contains two 
identical single-particle propagators and thus a factor of $1/2!$ must be 
introduced.

%



\section{Virial coefficients of the resonant Fermi gas
\label{sec:vcs}}

After the quite general discussion of the virial expansion above, we
now confine ourselves to the specific case of a spin- and
mass-balanced Fermi gas in three dimensions, described by the
Hamiltonian \eqref{eq:2ch}~\footnote{
For the virial expansion in two dimensions, see for instance 
Refs.~\cite{Liu2010,Ngamp2013b}.
}.  We present the second, third, and fourth
virial coefficients for the resonant Fermi gas, calculated using the
technique described in the preceding section. Our calculation of the
fourth virial coefficient is perturbative for a narrow resonance where
$T\gg1/mR^{*2}$, and approximate in the broad resonance case. All
other results are numerically exact.

\subsection{Second virial coefficient}

\begin{figure}
\centering
\includegraphics[width=0.866\linewidth]{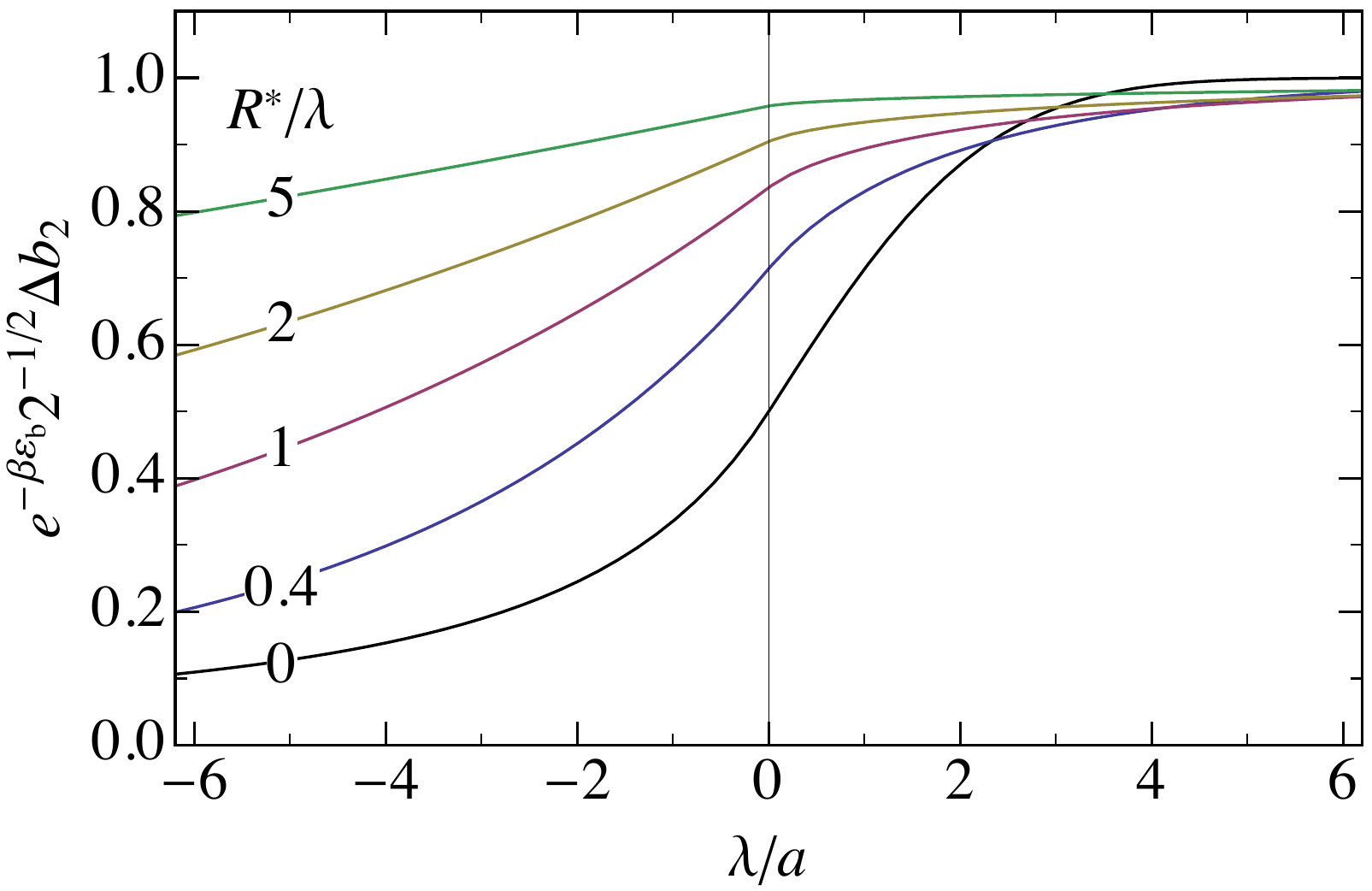}
\caption{The interaction part of the second virial coefficient as a
  function of interaction strength for several values of
  $R^*/\lambda$. As $\lambda/a\to\infty$ the virial coefficient
  approaches $\sqrt2e^{\beta\eb}$ and we normalize by this factor
  (taking $\eb=0$ for $a\leq0$).}
\label{fig:b2broad}
\end{figure}

In the spin- and mass-balanced gas, the second virial coefficient
takes the value $-2^{-5/2}$ in the absence of interactions --- see
Eq.~\eqref{eq:bfree}. The part of the second virial coefficient
arising from interactions can be obtained by inserting the two-body
$T$ matrix (\ref{2body-Tmat}) in the expression for the second virial
coefficient, Eq.~(\ref{eq:B11}). This yields
\begin{align}
  \Delta b_2 = \sqrt 2\oint'_E
  \frac{mR^*+\frac{m}{2\sqrt{-mE}}}{a^{-1}-\sqrt{-mE}+mR^*E}.
\end{align}
As depicted in Fig.~\ref{fig:b2broad}, $\Delta b_2$ increases
monotonically from the Fermi regime, through unitarity, to the Bose
regime. Consider first the limits $|a|\ll \lambda$ where the
temperature is much smaller than the energy scale of few-body physics.
Owing to the Boltzmann factor $\exp[-\beta E]$, the leading
contribution of the contour integral over $E$ then comes from an
interval of length $\sim T$ starting at the leftmost point of the
non-analytic structure of the $T$ matrix. Using this to evaluate the
virial coefficient in the Fermi limit $\lambda/a \ll -1$, the asymptotic
form is $\Delta b_2=-a/\lambda + 3\pi R^*a^2/\lambda^3$ for $|a| \ll R^*\lesssim\lambda$. 
On the other hand, in the Bose limit $\lambda/a \gg 1$, we have $\Delta
b_2=\sqrt2e^{\beta\eb}-a/\lambda$ for $R^*\lesssim\lambda$. Here, the
leading term may be understood as arising from the first term of the
virial expansion of a non-interacting Bose gas.  Indeed, we identify
the contribution to the grand potential \eqref{eq:grand} from the two
interacting fermions with that of a single boson
\begin{align}
-\frac1\beta\frac{2V}{\lambda^3}\Delta b_2 z^2
& =-\frac1\beta\frac V{\lambda_{\text{Bose}}^3}
 b_1^{\text{Bose}}z_{\text{Bose}}.
\label{eq:b2bose}
\end{align}
The factor two on the left hand side is due to the two fermionic
species. Using the first virial coefficient of the ideal Bose gas,
$b_1^{\text{Bose}}=1$, as well as the relation between the bosonic and fermionic
chemical potentials $\mu_{\text{Bose}}=2\mu+\eb$, the fugacity
$z_{\text{Bose}}=e^{\beta\mu_{\text{Bose}}}$, and the thermal wavelength
$\lambda_{\text{Bose}}=\lambda/\sqrt2$, we arrive at the asymptotic
form of $\Delta b_2$. We emphasize that this form of the virial
coefficient implies that the system is a non-interacting Bose gas of
dimers.

We next consider the evolution of $\Delta b_2$ at unitarity, as a
function of $R^*/\lambda$. For broad resonances (where $R^*$ may be
taken to vanish), or for temperatures $T\ll T_{R^*}\equiv1/mR^{*2}$,
the virial coefficient takes the well-known value $1/\sqrt2$. On the
other hand, for temperatures $T\gg T_{R^*}$ the virial coefficient is
approximately $\Delta b_2\approx \sqrt2- \frac{\lambda}{\pi
  R^*}$. This increase of the virial coefficient from $1/\sqrt2$ to
$\sqrt2$ as temperature increases above the scale set by $T_{R^*}$ has
been taken to imply that the two-particle system becomes more strongly
interacting~\cite{Ho2012,Peng2014}.  We, however, argue the opposite,
namely that the system approaches a non-interacting limit. To
understand this, it is convenient to introduce the pair propagator
\begin{align}
  D(E)\equiv T_2(E)/g^2 = \frac{1}{E+ \frac{m g^2}{4\pi}
    \left(-\sqrt{-mE}+a^{-1}\right)}.
\label{eq:pairprop}
\end{align}
At unitarity, this has the limiting behavior
\begin{align}
D(E)
\sim &
 \left\{ 
  \begin{array}{l l}
    R^*E^{-1/2} &\quad\text{if}\quad E\sim T \ll T_{R^*} \\
    E^{-1} &\quad\text{if}\quad E\sim T \gg T_{R^*} 
  \end{array} \right. \nn
\end{align}
That is, the pair propagator evolves from a resonance in the
low-temperature limit into a free propagator of a zero-energy 
  state as temperature increases beyond the energy scale set by
the effective range. Thus, the two fermions simply populate this
(non-interacting) pair state which, 
 as discussed in the
above paragraph, has virial coefficient $\sqrt2$. Indeed, the above
discussion carries over to the Fermi regime where, as long as $|a|\gg
R^*$, the virial coefficient becomes that of the non-interacting pair,
approaching $\sqrt2$ as seen in Fig.~\ref{fig:b2broad}.

\subsection{Third virial coefficient}

In the absence of interactions, the third virial coefficient in the
spin- and mass-balanced gas takes the value $3^{-5/2}$; see
Eq.~\eqref{eq:bfree}. We now proceed to calculate the effect of
interactions, which is encapsulated in $\Delta b_3=B_{21}$, with
$B_{21}$ defined in Eq.~\eqref{eq:B21}. The calculation now requires
one to solve the three-body problem in full, which has been the
subject of several works in the literature. Here we mainly follow
Refs.~\cite{Bedaque1998,Levinsen2011}, but see also Ref.~\cite{Reiner1966}
for a related framework for the calculation of the third virial coefficient. 
For completeness, we include a
discussion of the three-body problem in Appendix \ref{app:STM}.

\begin{figure}
\centering
\includegraphics[width=0.891\linewidth]{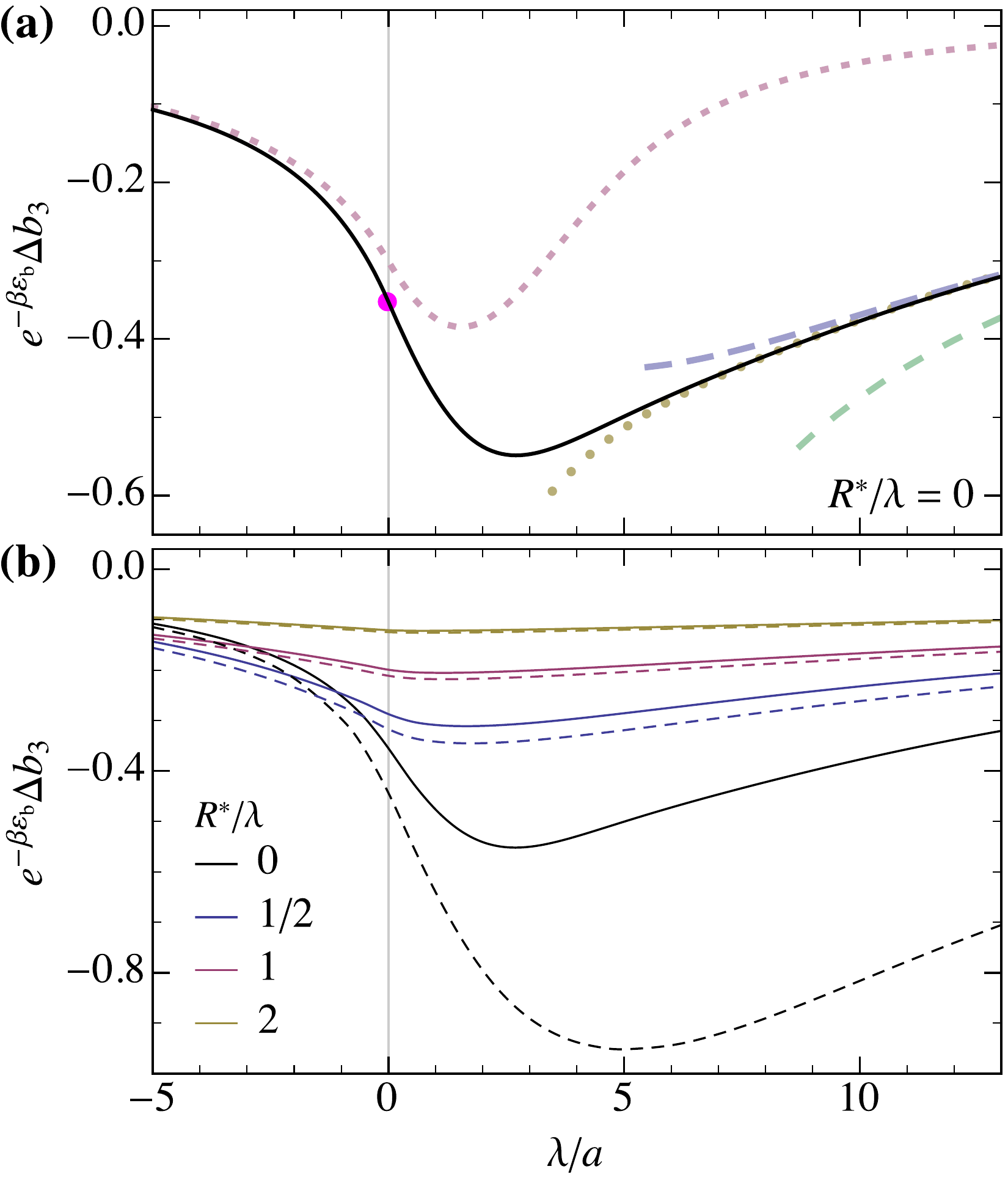}
\caption{The interaction part of the third virial coefficient as a
  function of interaction parameter. In the Bose regime the virial
  coefficient is proportional to $e^{\beta\eb}$ --- see
  Eq.~\eqref{eq:BU3body} --- and we normalize by this factor (taking
  $\eb=0$ for $a\leq0$). (a) In the broad resonance limit,
  $R^*/\lambda=0$, we compare $\Delta b_3$ (solid) with the two-body
  contribution which dominates in the Fermi regime (short dashed), and
  the results of the Beth-Uhlenbeck formula~\eqref{eq:BU3body}
  including $s$-wave only (dot-dashed), $s$- and $p$-wave (long
  dashed), and up to $d$-wave (dotted). The circle at $\lambda/a=0$
  depicts the experimentally determined virial coefficient from
  Refs.~\cite{Salomon2010,Zwierlein2012}. (b) The effect of a finite
  range parameter on $\Delta b_3$ (solid), and the approximation given
  by including only the diagrams of Eqs.~(\ref{b3-ad-perp}-c) (dashed).}
  \label{fig:b3}
\end{figure}

Consider the crossover from the Fermi to the Bose regimes, as depicted
in Fig.~\ref{fig:b3}. In the broad resonance case shown in
Fig.~\ref{fig:b3}(a), our results completely match those of
Refs.~\cite{Liu2009,Leyronas2011} in the whole range of interactions,
and Refs.~\cite{Kaplan2011,Rakshit2012} at unitarity. The results do not
match those of Ref.~\cite{Rupak2007}, in which the three-body problem
was confined to the $s$-wave channel. We first discuss the asymptotic
limits in which the few-body energy scale set by the two-body
scattering length is much greater than temperature, focusing on the
broad resonance case for simplicity. Then, we have the scaling of the
three contributions of Eq.~\eqref{eq:B21} to the interaction part of
the virial coefficient:
%
\begin{subequations}
\begin{align}
\label{b3scaling:atomdimer}
\frac{\lambda^3}{V}\oint'_E
\begin{tikzpicture}[baseline={([yshift=-.6ex]current  bounding  box.center)}]
\draw[blue] (0.2,-0.1) -- (0,-0.1);
\draw[red] (0,0.1) -- (0.2,-0.1);
\draw[T2] (0.2,-0.1) -- (0.4,-0.1); \draw[blue] (0,0.3) -- (0.4,0.3);
\fill[gray,opacity=0.5] (0.4,-0.1) rectangle (0.8,0.3) ;
\node at (0.6,0.1) {$t_3$};
\draw[blue] (0.8,-0.1) -- (1.2,-0.1) ;
\draw[T2] (0.8,0.3) -- (1,0.3);
\draw[blue] (1.2,0.3) -- (1,0.3);\draw[red] (1,0.3) -- (1.2,0.1);
\node[blue] at (-0.1,-0.1) {1};\node at (-0.1,0.1) [red]{1};\node[blue] at (-0.1,0.3) {2};
\node[blue] at (1.3,-0.1) {2};\node at (1.3,0.1) [red]{1};\node[blue] at (1.3,0.3) {1};
\end{tikzpicture}
 \sim &
 \left\{ 
  \begin{array}{l l}
    (\frac{a}{\lambda})e^{\beta\eb} &\text{Bose} \\
    (\frac{a}{\lambda})^2 &\text{Fermi}
  \end{array} \right.
\\ \label{b3scaling:2body}
\frac{\lambda^3}{V}\oint'_E
\begin{tikzpicture}[baseline={([yshift=-.6ex]current  bounding  box.center)}]
\draw[red] (0,0.1) -- (0.2,-0.1);\draw[blue] (0.2,-0.1) -- (0,-0.1);
\draw[red] (0.6,0.1) -- (0.4,-0.1);\draw[blue] (0.4,-0.1) -- (0.6,-0.1);
\draw[T2] (0.2,-0.1) -- (0.4,-0.1); \draw[blue] (0,0.3) -- (0.6,0.3);
\node[blue] at (-0.1,-0.1) {1};\node at (-0.1,0.1) [red]{1};\node[blue] at (-0.1,0.3) {2};
\node[blue] at (0.7,-0.1) {2};\node at (0.7,0.1) [red]{1};\node[blue] at (0.7,0.3) {1};
\end{tikzpicture}
 \sim &
 \left\{ 
  \begin{array}{l l}
    (\frac{a}{\lambda})^3e^{\beta\eb} &\text{Bose} \\
    \frac{a}{\lambda} &\text{Fermi}
  \end{array} \right.
\\
\frac{\lambda^3}{V}\oint'_E
\begin{tikzpicture}[baseline={([yshift=-.6ex]current  bounding  box.center)}]
\draw[blue] (0.2,-0.1) -- (0,-0.1);
\draw[red] (0,0.1) -- (0.2,-0.1);
\draw[T2] (0.2,-0.1) -- (0.4,-0.1); \draw[blue] (0,0.3) -- (0.4,0.3);
\fill[gray,opacity=0.5] (0.4,-0.1) rectangle (0.8,0.3) ;
\node at (0.6,0.1) {$t_3$};
\draw[blue] (0.8,-0.1) -- (1.2,-0.1) ;
\draw[T2] (0.8,0.3) -- (1,0.3);
\draw[blue] (1.2,0.3) -- (1,0.3);\draw[red] (1,0.3) -- (1.2,0.1);
\node[blue] at (-0.1,-0.1) {1};\node at (-0.1,0.1) [red]{1};\node[blue] at (-0.1,0.3) {2};
\node[blue] at (1.3,-0.1) {1};\node at (1.3,0.1) [red]{1};\node[blue] at (1.3,0.3) {2};
\end{tikzpicture}
\sim &
 \left\{ 
  \begin{array}{l l}
    (\frac{a}{\lambda})^3e^{\beta\eb} &\text{Bose} \\
    (\frac{a}{\lambda})^2 &\text{Fermi}
  \end{array} \right.
\label{eq:b3last}
\end{align}
\end{subequations}
%
where the explicit expressions for the diagrams is clear from
Eq.~\eqref{eq:B21}. In the Fermi limit, the scaling follows from all
momenta and energies being set by $\lambda$. Therefore, $T_2\sim a/m$
and $t_3\sim m\lambda^2$. On the other hand, in the Bose limit, 
the Boltzmann factor in the 
two-body $T$ matrix forces the energy $E=-\eb+E_\text{col}$ to be within a
collision energy $E_\text{col}\sim1/m\lambda^2$ of the binding
energy. Thus $T_2(-\eb+E_\text{col})\approx
\frac{8\pi}{m^2a}\frac{1}{E_\text{col}}\sim\lambda^2/ma$ and $t_3\sim
ma^2$. In Eq.~(\ref{eq:b3last}), due to the exchange process, two
momenta are integrated over, and only one of these is suppressed by
the Boltzmann factor; consequently, in this diagram the momenta count
as $1/\lambda$ and $1/a$, respectively.

\begin{figure}
\centering
\includegraphics[width=0.879\linewidth]{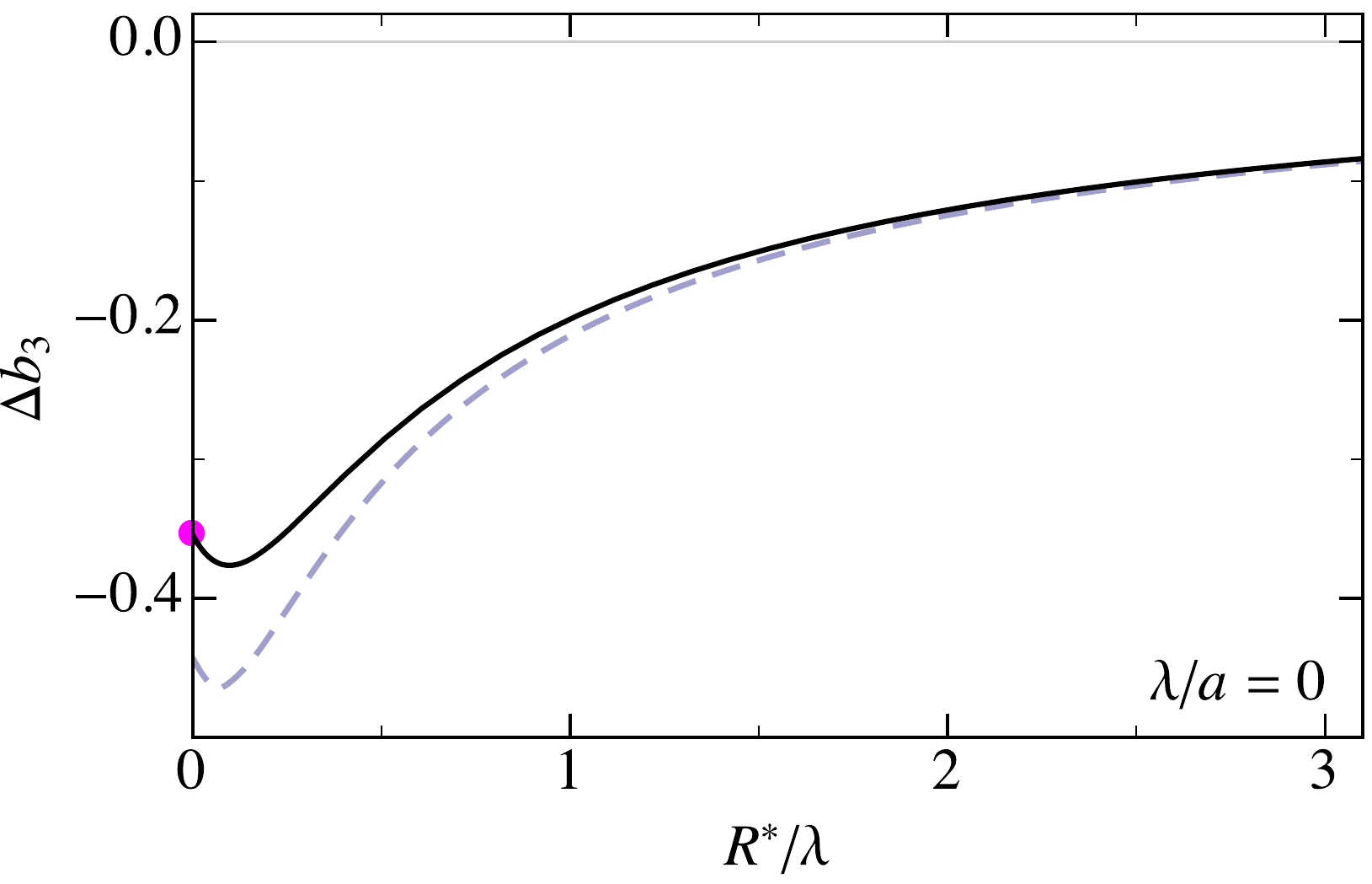}
\caption{The interaction part of the third virial coefficient (solid
  line) at unitarity as a function of $R^*/\lambda$. The dashed line
  depicts the result of considering only the diagrams of
  Eqs.~(\ref{b3-ad-perp}-c), corresponding to replacing the full
  three-body $T$ matrix by its Born approximation. The dotted curve is
  the narrow resonance asymptote --- see Eq.~\eqref{eq:b3asymp}.}
  \label{fig:b3narrow}
\end{figure}

As may be anticipated, the two-body contribution,
Eq.~(\ref{b3scaling:2body}), is the dominant contribution in the
weakly interacting Fermi regime. On the other hand, in the Bose limit,
Eq.~(\ref{b3scaling:atomdimer}) is the leading contribution. This has
a very natural interpretation: Due to the lack of exchange of the
external legs, this diagram corresponds to atom-dimer scattering, and
thus to an effective two-body process. Then we may recast the virial
coefficient in Beth-Uhlenbeck form, similarly to the two-body problem in
Eq.~(\ref{eq:B11}). The result is
\begin{align}
\Delta b_3 \overset{\lambda\gg a}{\approx}
 3^{\frac{3}{2}}e^{\beta\eb} \sum_{\ell}\xi_\ell\int_0^\infty
\frac{dk}{\pi}e^{-\beta \frac{3k^2}{4m}}
\delta'_{\text{ad},\ell}(k),
\label{eq:BU3body}
\end{align}
where $\delta_{\text{ad},\ell}$ is the atom-dimer phase shift in the
$\ell$'th partial wave. For details on the derivation of
Eq.~\eqref{eq:BU3body}, see Appendix \ref{app:BU3body}. If the
temperature is negligible compared with the binding energy, only the
low-energy behavior of the scattering phase shifts matter, i.e., the
atom-dimer scattering length $a_\text{ad}
=-\lim_{k\to0}k^{-1}\tan\delta_{\text{ad},\ell}(k)= 1.18a$ \cite{stm}. For
decreasing $\lambda/a$ we need to take the derivative of the phase
shifts obtained in Refs.~\cite{Levinsen2009,Levinsen2011}. The result
is shown in Fig.~\ref{fig:b3}(a) and yields a good approximation to
the virial coefficient for $\lambda/a\gtrsim5$.

Next we investigate the behavior of the virial coefficient for a finite
range parameter --- see Fig.~\ref{fig:b3}(b). First of all, we see that
increasing $R^*/\lambda$ tends to suppress the virial coefficient for
fixed $\lambda/a$. Secondly, we see that a perturbative approach works
very well already at $R^*/\lambda=1$. 
The particular approach we use is 
to take the first Born approximation of the three-body 
$T$ matrix, i.e., keeping only the first term in Eq.~(\ref{3bodyTmatrix}). 
That is, the three diagrams contributing to $\Delta b_3$ take the 
form:
\begin{subequations}
\begin{align}
\label{b3-ad-perp}\hspace{-1mm}
-
\begin{tikzpicture}[baseline={([yshift=-.6ex]current  bounding  box.center)}]
\draw[red] (0,0.1) -- (0.2,-0.1);\draw[blue] (0.2,-0.1) -- (0,-0.1);
\draw[T2] (0.2,-0.1) -- (0.4,-0.1); \draw[blue] (0,0.3) -- (0.6,0.3);
\draw[red] (0.6,0.3) -- (0.4,-0.1);\draw[blue] (0.4,-0.1) -- (1,-0.1);
\draw[T2] (0.6,0.3) -- (0.8,0.3);
\draw[blue] (1,0.3) -- (0.8,0.3);\draw[red] (0.8,0.3) -- (1,0.1);
\node[blue] at (-0.1,-0.1) {1};\node at (-0.1,0.1) [red]{1};\node[blue] at (-0.1,0.3) {2};
\node[blue] at (1.1,-0.1) {2};\node at (1.1,0.1) [red]{1};\node[blue] at (1.1,0.3) {1};
\end{tikzpicture}
&= 
-\sum_\p\frac{ T_2^2(E-\frac{3p^2}{4m})\frac{dT_2^{-1}(E-\frac{3p^2}{4m})}{dE}}
{E-\frac{3p^2}{m}},
\\
\label{b3-2b-perp}
-
\begin{tikzpicture}[baseline={([yshift=-.6ex]current  bounding  box.center)}]
\draw[red] (0,0.1) -- (0.2,-0.1);\draw[blue] (0.2,-0.1) -- (0,-0.1);
\draw[red] (0.6,0.1) -- (0.4,-0.1);\draw[blue] (0.4,-0.1) -- (0.6,-0.1);
\draw[T2] (0.2,-0.1) -- (0.4,-0.1); \draw[blue] (0,0.3) -- (0.6,0.3);
\node[blue] at (-0.1,-0.1) {1};\node at (-0.1,0.1) [red]{1};\node[blue] at (-0.1,0.3) {2};
\node[blue] at (0.7,-0.1) {2};\node at (0.7,0.1) [red]{1};\node[blue] at (0.7,0.3) {1};
\end{tikzpicture}
&= 
-\sum_\p \frac{T_2(E-\frac{3p^2}{4m})}{\left(E-\frac{3p^2}{m}\right)^2},
\\
\label{b3-3a-perp}
\begin{tikzpicture}[baseline={([yshift=-.6ex]current  bounding  box.center)}]
\draw[red] (0,0.1) -- (0.2,-0.1);\draw[blue] (0.2,-0.1) -- (0,-0.1);
\draw[T2] (0.2,-0.1) -- (0.4,-0.1); \draw[blue] (0,0.3) -- (0.6,0.3);
\draw[red] (0.6,0.3) -- (0.4,-0.1);\draw[blue] (0.4,-0.1) -- (1,-0.1);
\draw[T2] (0.6,0.3) -- (0.8,0.3);
\draw[blue] (1,0.3) -- (0.8,0.3);\draw[red] (0.8,0.3) -- (1,0.1);
\node[blue] at (-0.1,-0.1) {1};\node at (-0.1,0.1) [red]{1};\node[blue] at (-0.1,0.3) {2};
\node[blue] at (1.1,-0.1) {1};\node at (1.1,0.1) [red]{1};\node[blue] at (1.1,0.3) {2};
\end{tikzpicture}
&=\sum_{\p,\p'}
\frac{T_2(E-\frac{3p^2}{4m})T_2(E-\frac{3p'^2}{4m})}
{\left(E-\ep-\epsilon_{\p'}-\epsilon_{\p+\p'}\right)^3}.
\end{align}
\end{subequations}
The idea is that keeping the exact pole structure of the
two-body $T$ matrix may help control the approximation in the regime where
$R^*\sim\lambda$.  
The close agreement between the exact calculation and our
approximation observed in Fig.~\ref{fig:b3}(b) appears to validate our
approach. The main difference between the perturbative approach and
the exact takes place for a broad resonance in the Bose regime: The
source of the discrepancy is that for a broad resonance the Born
approximation of atom-dimer scattering predicts $a_\text{ad}=8a/3$,
while the exact calculation gives $a_\text{ad}=1.18a$ \cite{stm}, and
thus the atom-dimer interaction is overestimated in the Born
approximation. We further illustrate the comparison between the exact,
approximate (in the above sense), and truly perturbative (expanding
$T_2(E)$ in powers of $1/R^*$) approaches in
Fig.~\ref{fig:b3narrow}. It is seen that our approximation provides a
very good agreement with the exact virial coefficient, even as
$R^*/\lambda$ approaches zero. We find the exact value at $R^*=0$ to be 
$\Delta b_3\simeq -0.3551$, in agreement with 
Refs.~\cite{Liu2009,Leyronas2011,Rakshit2012}.

Finally, we note that $\Delta b_3$ at unitarity is non-monotonic as a
function of $R^*/\lambda$, a feature which is present in both the
exact and approximate results --- see Fig.~\ref{fig:b3narrow}. This
reflects how at small $R^*/\lambda$ a new interaction channel becomes
available, increasing the magnitude of the virial coefficient, whereas
at large $R^*/\lambda$ the interactions are suppressed. In the limit
$R^*/\lambda\gg1$ we evaluate the asymptotic form of the virial
coefficient:
\begin{align}
\Delta b_3 \overset{\lambda\ll R^*}{\approx}
  -\frac{2\sqrt{2}}{3\pi}\frac{\lambda}{R^*}+\frac{4 (5 \sqrt{3} \pi-18)}{27
    \pi^2}\left(\frac{\lambda}{R^*}\right)^2. 
\label{eq:b3asymp}
\end{align}
In Fig.~\ref{fig:b3narrow} we see that this asymptotic expression works
very well for $R^*/\lambda\gtrsim2$.

\subsection{Fourth virial coefficient}

In the absence of interactions, the fourth virial coefficient takes
the value: $b_4^\text{free}=-4^{-5/2}$. In order to obtain the
contribution to the virial coefficient arising from interactions, one
may in principle extend the above diagrammatic analysis; however the
task of solving the four-body problem exactly is numerically
taxing. 
Instead, we take a pragmatic approach and consider
the Born approximation of the four-body problem, summing all diagrams
containing at most four two-body propagators. 
This includes the first Born approximation of the dimer-dimer scattering 
$T$ matrix. 
The relevant diagrams are shown in
subsections \ref{sec:b22} and \ref{sec:b31}. This is the same
approximation which was shown above to work very well for $\Delta b_3$
once $R^*/\lambda\geq1$; at unitarity, it even gave a reasonable result
once $R^*\to0$. The two- and three-body contributions to $\Delta b_4$
are computed exactly.

\begin{figure}
\centering
\includegraphics[width=0.866\linewidth]{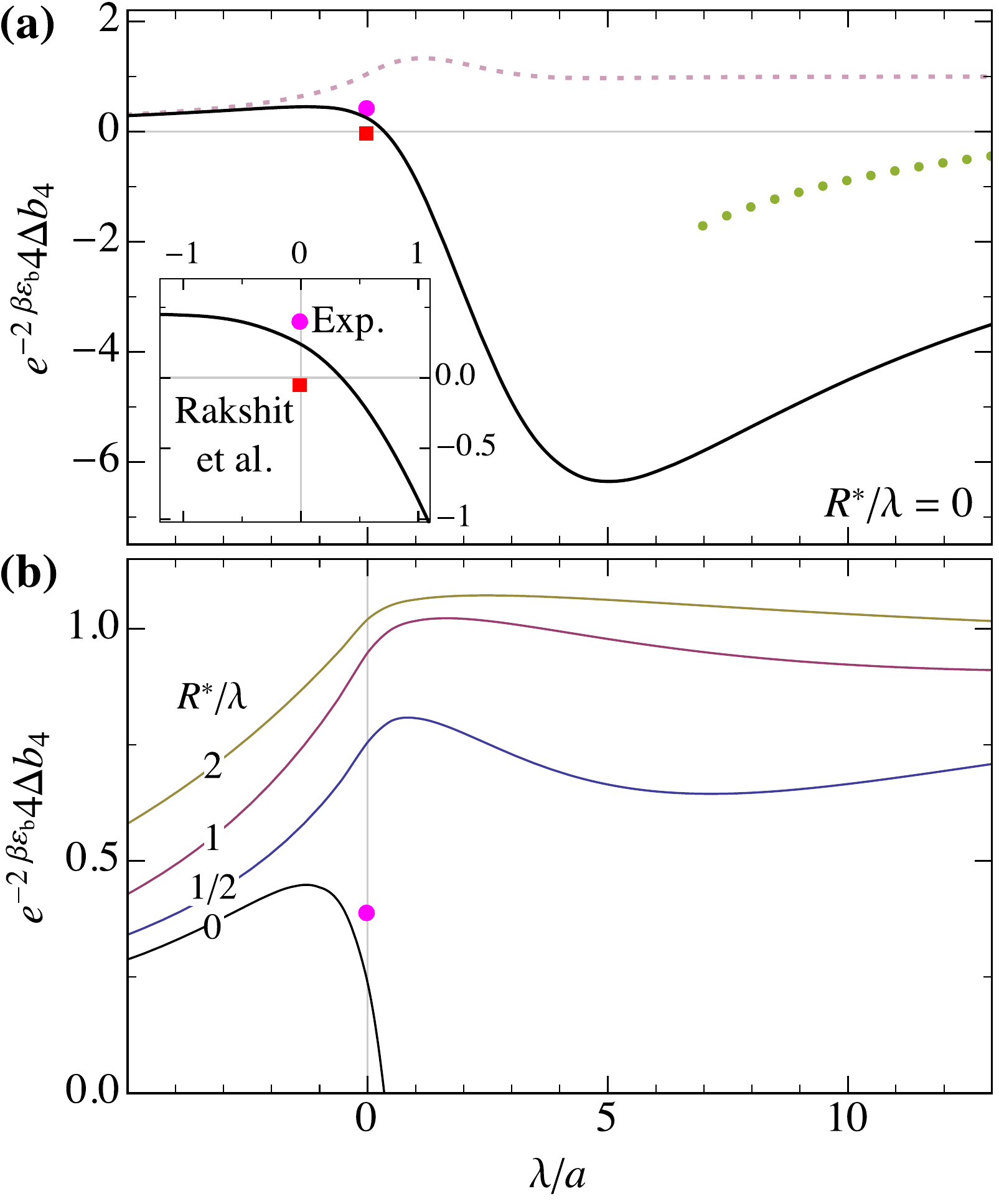}
\caption{The interaction part of the fourth virial coefficient as a
  function of the interaction parameter $\lambda/a$. In the Bose regime,
  $\Delta b_4\to e^{2\beta\eb}/4$ and we normalize by this factor,
  taking $\eb=0$ for $a\leq0$. (a) For a broad resonance with
  $R^*/\lambda=0$ we display our approximate value of $\Delta b_4$
  (solid), along with the dominant two-body contribution in the Fermi
  regime (short dashed), and the dominant contribution in the Bose limit,
  Eq.~\eqref{eq:ddBU}, taking 
  $a_\text{dd} \approx 0.6a$ (dotted). The circle
  represents the experimental measurement \cite{Salomon2010,
    Zwierlein2012} and the square the calculation of
  Ref.~\cite{Rakshit2012}. (b) Taking the effective range into account,
  we display the virial coefficient for $R^*/\lambda=0$, 1/2, 1, 2
  (bottom to top).}
\label{fig:b4broad}
\end{figure}

Figure \ref{fig:b4broad}(a) depicts the virial coefficient within our
approximation for a broad resonance with $R^*/\lambda=0$. We see that
in the weakly interacting Fermi limit of $\lambda/a\ll-1$, the two-body
contribution dominates. On the other hand, in the Bose limit where
$\lambda/a\gg1$, the four fermions may be approximated by two deeply
bound dimers. The virial coefficient is then a sum of a contribution
arising from the second-order term of a non-interacting Bose gas, and
one from the dimer-dimer interaction. Employing for the former the
same arguments which related the second virial coefficient of the
Fermi gas with the first of a non-interacting Bose gas, see
Eq.~\eqref{eq:b2bose}, and for the latter the same arguments which
lead to the Beth-Uhlenbeck expression for the third virial coefficient
in the Bose regime, Eq.~\eqref{eq:BU3body}, we find
\begin{align}\nn
  \Delta b_{4}&\overset{\lambda\gg a}{\approx}
  e^{2\beta\eb}\!\!
\left[\frac{1}{4}+8\sum_\ell\xi_\ell\int_0^\infty
    \frac{dk}{\pi}e^{-\beta\frac{k^2}{2m}}\delta_{\text{dd},\ell}'(k)\right]
\!
\\
&\overset{\phantom{\lambda\gg a}}{\approx}
 e^{2\beta\eb}
\left(\frac{1}{4}-8\frac{a_\text{dd}}{\lambda}\right),
\label{eq:ddBU}
\end{align}
where 
we approximate $-k^{-1}\tan\delta_{\text{dd},0}(k)\approx a_\text{dd}$ and 
ignore all phase shifts other than $s$-wave. 
The correct dimer-dimer scattering length is $a_\text{dd}=0.6a$, 
as derived in Ref.~\cite{Petrov2004}; for a diagrammatic derivation 
closer to the present formulation, see 
Refs.~\cite{brodsky2006,Levinsen2006}. 
We see that the asymptote \eqref{eq:ddBU} gives a much smaller dimer-dimer
interaction shift of the virial coefficient compared to our approximate result. This is because our approach
contains only the leading diagram of the Born approximation of 
dimer-dimer scattering, which overestimates the dimer-dimer repulsion
and yields $a_\text{dd}=2a$. 
Nevertheless, we expect our result to recover the qualitative behavior 
of $\Delta b_4$ across the crossover. 
In particular, we see that the coefficient is positive in the Fermi regime 
and changes sign close to unitarity mainly due to the dimer-dimer repulsion 
at large positive $\lambda/a$. 
Finally, it changes sign again and becomes positive --- using $a_\text{dd}=0.6a$ in
Eq.~\eqref{eq:ddBU} we estimate this to occur
for $\lambda/a\approx19$.

Increasing the range parameter for fixed scattering length suppresses
the dimer-dimer interaction --- see
Ref.~\cite{Levinsen2011}. Consequently, as shown in
Fig.~\ref{fig:b4broad}(b), already at $R^*/\lambda=1/2$ we do not
expect that the virial coefficient changes sign in the crossover. As
in the three-body case, we expect the virial coefficient calculated
within our approximation to work well once $R^*/\lambda\geq1$.

\begin{figure}
\centering
\includegraphics[width=0.851\linewidth]{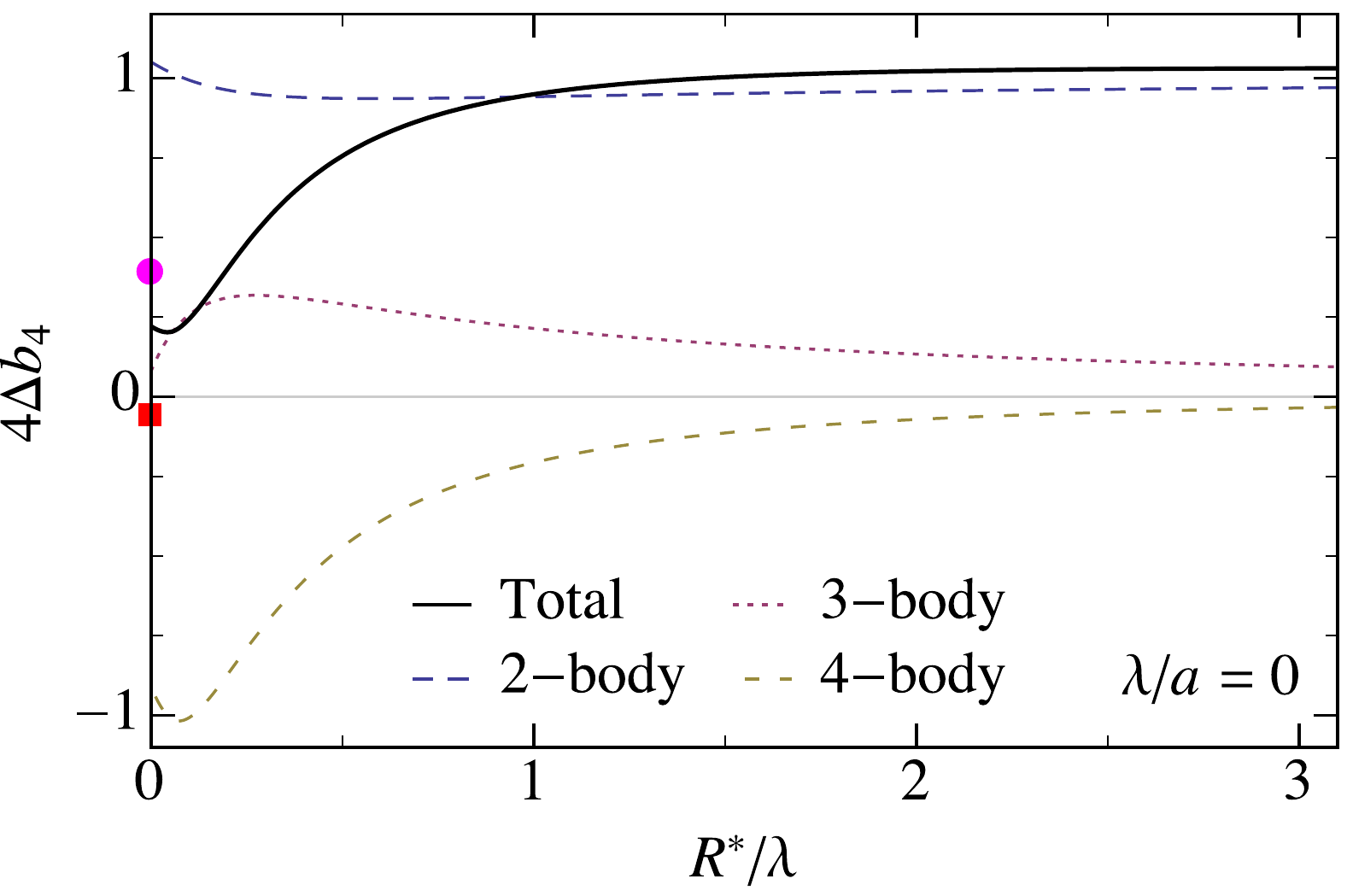}
\caption{The interaction part of the fourth virial coefficient at
  unitarity ($1/a=0$) as a function of the effective range
  $R^*/\lambda$. The four-body contribution to the virial coefficient
  is approximated as explained in the text.}
\label{fig:b4narrow}
\end{figure}

Consider finally the unitary Fermi gas, where we find $\Delta
b_4\approx0.06$ for a broad resonance --- see
Fig.~\ref{fig:b4narrow}. This is in contrast to the previous
theoretical work of Ref.~\cite{Rakshit2012}, where $\Delta b_4$ was
found to be negative \footnote{Note that the authors of
  Ref.~\cite{Rakshit2012} recognised that their calculation did not
  predict the correct fourth virial coefficient.}. We note however
that our $\Delta b_4$ is still smaller than the experimental
value $\Delta b_4^\text{exp.}\approx 0.096$ \cite{Salomon2010,
  Zwierlein2012}, most likely due to the overestimation of the
dimer-dimer repulsion within our approximation. Our results also
illustrate the difficulty in determining the fourth virial
coefficient, as the coefficient is expected to show a strong
non-monotonic behavior in the vicinity of the resonance. For large
$R^*/\lambda$ we see that the two-body contribution to the virial
coefficient dominates and $\Delta b_4\to1/4$. This is again an
illustration of how the system evolves towards a non-interacting Bose
gas in this limit. The first correction to this result arises from
both two- and the three-body contributions, and we find the limiting
behavior
\begin{align}
  \Delta b_4\overset{\lambda\ll R^*}{\approx} \frac14+\frac{9-4\sqrt3}{12\pi}\frac\lambda{R^*},
\end{align} 
for a narrow resonance at unitarity.

\subsection{Higher virial coefficients in the Fermi and Bose regimes}

It is clear from the preceding discussion that the interaction part of
the virial coefficient in the weakly interacting Fermi regime is
dominated by the two-body contribution, i.e., for $\lambda/a\ll-1$ we
have $\Delta b_N \approx \Delta b^\text{2-body}_N$. For reference, we
give this here up to arbitrary order
\begin{align}\label{eq:bN2body}
\Delta b^\text{2-body}_N =& \frac{(-1)^N}{N^{5/2}}
\oint'_E \frac{\frac{m}{\pi}T_2(E)}{\sqrt{-mE}}\sum_{j=0}^{N-2}(N-j-1)
\nn \\ & \hspace{-14mm}\times
\left[
e^{\zeta_N}(1+2\zeta_N)
\left(1+\text{Erf}\left[-\sqrt{\zeta_N}\right]\right)-\frac{2\sqrt{\zeta_N}}{\sqrt\pi}
\right],
\end{align}
where $\zeta_N=-\beta E[N-2-\frac{1}{N}(N-2j-2)^2]/2$.

In the opposite limit where $\lambda/a\gg1$, the virial coefficients
$\Delta b_N$ are related to those of a non-interacting Bose
gas. Specifically, as $\lambda/a\to\infty$ the virial coefficients
approach
\begin{align}
e^{-\frac{N}{2}\beta\eb}\Delta b_N &\rightarrow
\sqrt2 \left\{ 
  \begin{array}{l l}
    b_{N/2}^\text{Bose} = (N/2)^{-5/2} 
    &\quad\text{$N$ even} \\
    0  &\quad\text{$N$ odd}
  \end{array} \right. .
\label{eq:bNbose}
\end{align}
This result holds even at unitarity when $R^*/\lambda\gg1$. For
even $N$, this formula follows from comparing the contributions to the
grand potential from $N$ interacting fermions with that of $N/2$
non-interacting bosons, as in the case of the second virial
coefficient --- see Eq.~\eqref{eq:b2bose}.

The asymptotic form of the virial coefficients in the Bose limit where
the virial coefficients grow exponentially, Eq.~\eqref{eq:bNbose},
prompts us to define the effective fugacity
\begin{align}\label{eq:zstar}
z^* = \left\{ 
  \begin{array}{l l}
    z  &\quad\text{if}\quad a\le0 \\
    e^{\beta\eb/2}z  &\quad\text{if}\quad a>0
  \end{array} \right. ,
\end{align}
such that all coefficients in the power series in $z^*$ remain of
order unity throughout the crossover.



\section{Thermodynamics \label{sec:thermodyn}}

Having calculated the virial coefficients, we now turn to the
investigation of the thermodynamics of the resonant Fermi gas. In the
grand canonical ensemble, the thermodynamic variables of a homogeneous
system are functions of the chemical potential, temperature, and
volume. That is, the grand potential is $\Omega = \Omega(\mu,T,V)$
from which all thermodynamic variables can be obtained. The pressure
is given by
\begin{align}\label{eq:pressurevirial}
P=-\frac{\O}{V} & 
=P_0+2T\lambda^{-3}\sum_{N\ge2}\Delta b_{N}z^N,
\end{align}
where the ideal Fermi gas pressure is given by
$P_0=2T\lambda^{-3}[-\text{Li}_{5/2}(-z)]$, with $\text{Li}_\nu$ the
polylogarithm. The density reads
\begin{align}\label{eq:densityvirial}
n=-\left.\frac{\partial}{\partial\mu}\frac{\O}{V} \right|_{T,V}
&= n_0+\frac{2}{\lambda^3}\sum_{N\ge2}N\Delta b_{N}z^N,
\end{align}
where $n_0=2\lambda^{-3}[-\text{Li}_{3/2}(-z)]$ is the density of an
ideal Fermi gas.  Finally, we write down the entropy density $s\equiv
S/V$:
\begin{align}\label{eq:entropyvirial}
s=-\left.\frac{\partial}{\partial T}\frac{\O}{V} \right|_{\mu,V}
=\frac{5}{2}\frac{P}{T} -\frac{\mu}{T} n
+\frac{2}{\lambda^3}\sum_{N\ge2} T\frac{\partial \Delta
  b_{N}}{\partial T}z^N.
\end{align}
Note that we only have $\partial \Delta b_{N}/\partial T$ appearing in $s$ since 
the ideal Fermi gas virial coefficients do not depend on temperature.

\begin{figure}
\centering
\includegraphics[width=.859\linewidth]{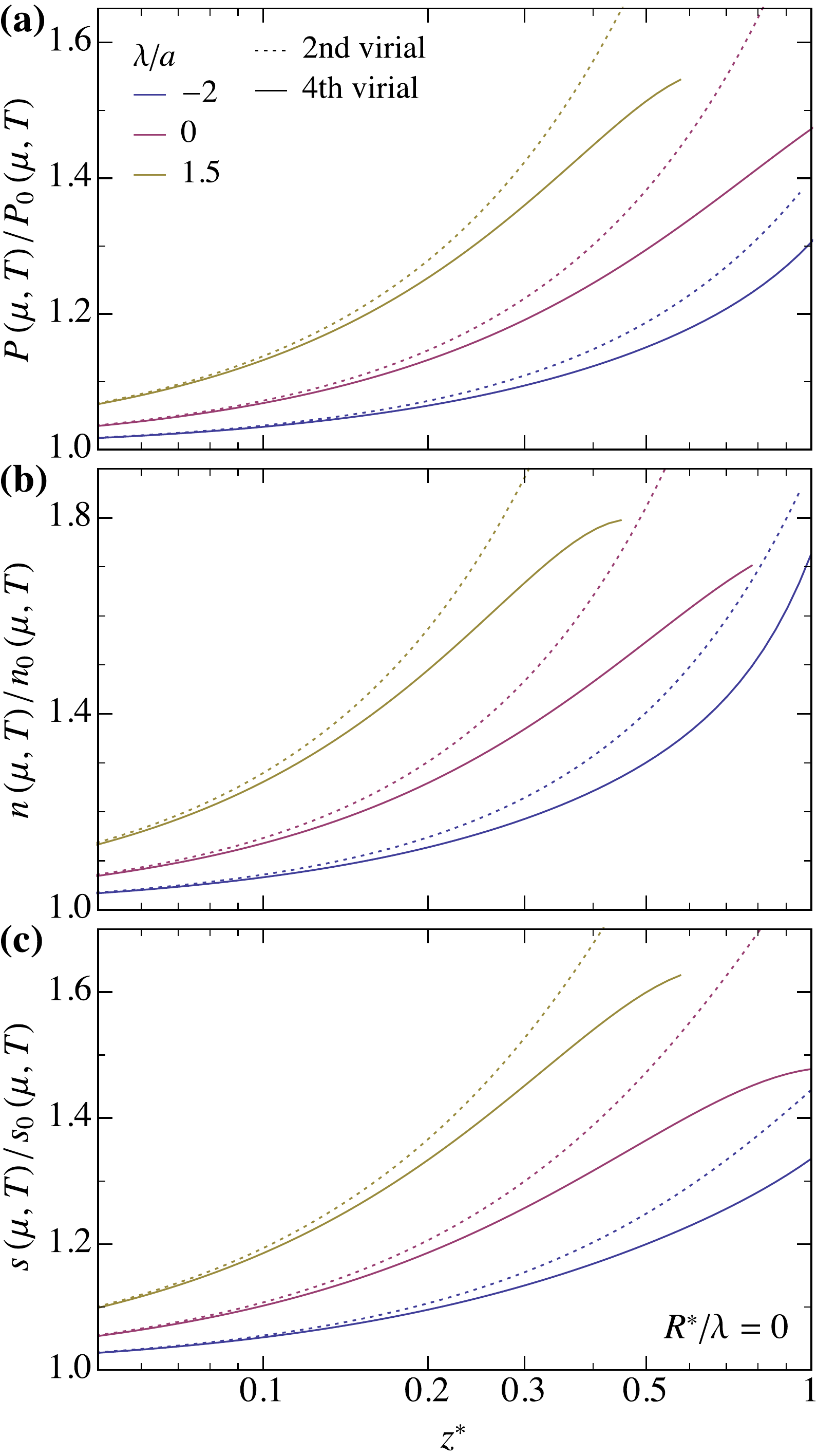}
\caption{The pressure $P$, density $n$, and entropy density $s$ in the grand canonical
  ensemble as a function of expansion parameter $z^*$ for a broad
  resonance $R^*=0$. We compare the results from the virial
  expansion at second and fourth order in the fugacity, for three
  different values of the interaction parameter $\lambda/a$: From top
  to bottom we consider the Bose regime with $\lambda/a=1.5$,
  unitarity, and the Fermi regime with $\lambda/a=-2$. For the latter,
  the virial coefficients are dominated by the two-body contribution
  \eqref{eq:bN2body}, the result of which is shown as a thick
  dashed line.}
  \label{fig:crossgrand}
\end{figure}

In Fig.~\ref{fig:crossgrand} we display the pressure, density, and
entropy for a broad resonance in the Fermi regime, at unitarity, and
in the Bose regime. We see from the comparison between the second and
fourth-order results that the virial expansion breaks down as $z^*$
approaches $1$, as expected. In the Fermi regime where
$\lambda/a\ll-1$, the virial coefficients are dominated by those of
the ideal Fermi gas, the leading correction due to interactions
arising from the two-body contributions \eqref{eq:bN2body}. In the
figure, this approximation is seen to give a clear improvement
compared with the second-order virial expansion. On the other hand, in
the Bose regime where $\lambda/a\gg1$, the virial coefficients are
dominated by those arising from the non-interacting gas of dimers,
Eq.~\eqref{eq:bNbose}, and we have
\begin{subequations}
\begin{align}
P &\approx 2\sqrt2T\lambda^{-3}\text{Li}_{5/2}(z^{*2}),
\label{eq:pbose}\\
n &\approx 4\sqrt2\lambda^{-3}\text{Li}_{3/2}(z^{*2}),
\label{eq:nbose}\\
s &\approx \sqrt2\lambda^{-3}\left[5\text{Li}_{5/2}(z^{*2})
-2\beta(2\mu+\eb)\text{Li}_{3/2}(z^{*2})\right].
\label{eq:sbose}
\end{align}
\end{subequations}
The leading correction arises from the dimer-dimer scattering length
contribution to $\Delta b_4$ --- see Eq.~\eqref{eq:ddBU}.  We do not
display the Bose limit expressions in the figure, as we see from
Fig.~\ref{fig:b4broad}(a) that these are not expected to be accurate
for the broad resonance until  larger $\lambda/a$.

We see in Fig.~\ref{fig:crossgrand}(a) that, contrary to what one
might naively expect, the pressure appears to increase as we go from
the Fermi to the Bose regime. The point is that this happens at a
fixed expansion parameter, causing the density to increase
simultaneously, as seen in Fig.~\ref{fig:crossgrand}(b). This effect
may also be understood as arising from the fact that the grand potential $\Omega$ decreases
from the Fermi to the Bose regime, due to the formation of pairs.

\begin{figure}
\centering
\includegraphics[width=.859\linewidth]{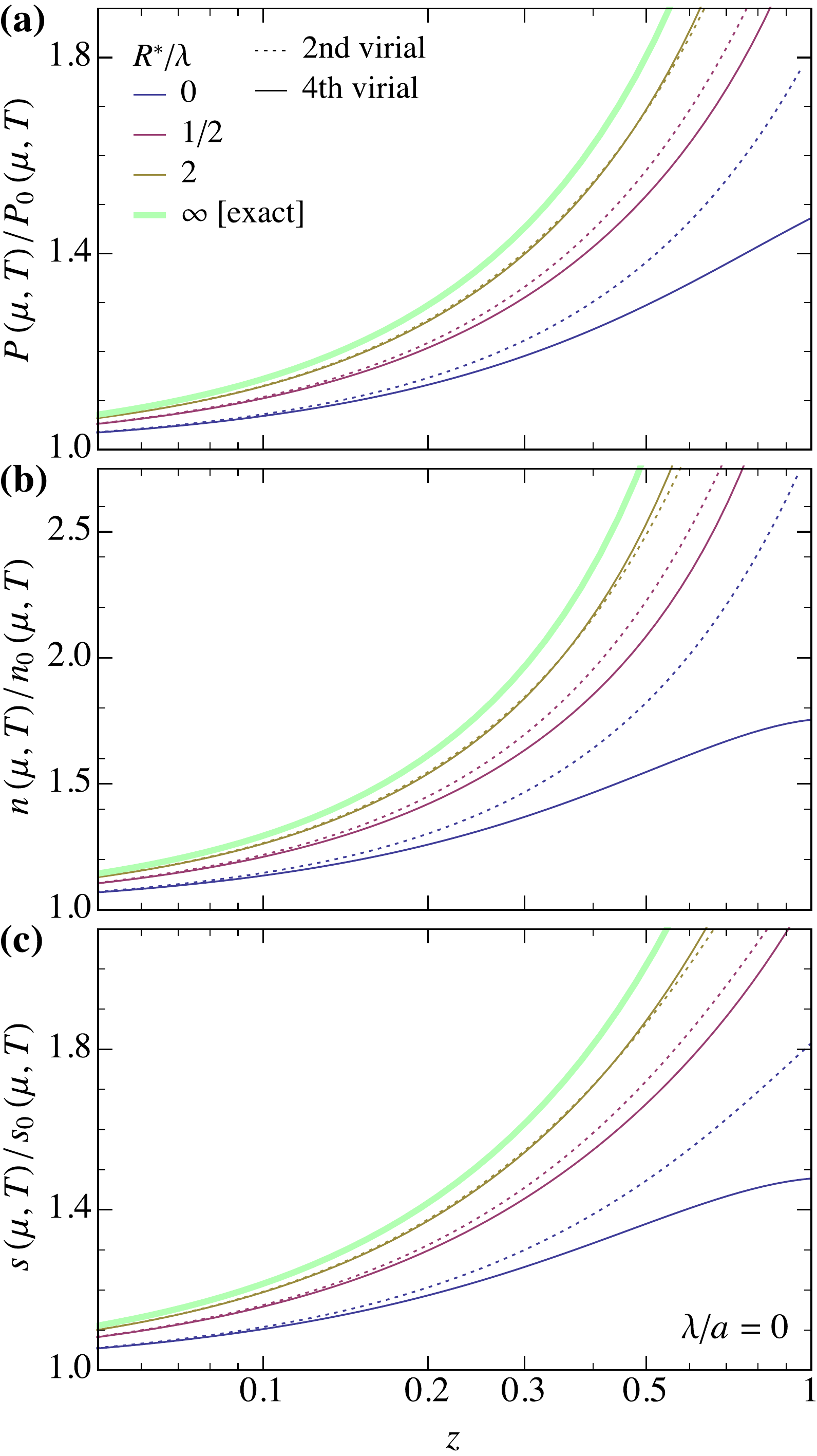}
\caption{The pressure, density, and entropy at resonance.  We compare
  the second and fourth orders of the virial expansion for various
  resonance widths. In the narrow resonance limit, the limiting
  behavior of the thermodynamic variables is given in
  Eqs.~(\ref{eq:pbose}-c).  }
\label{fig:thermounitary}
\end{figure}

In Fig.~\ref{fig:thermounitary}, we consider the behavior of the
thermodynamic variables at unitarity as a function of the range
parameter, $R^*/\lambda$. As discussed in Sec.~\ref{sec:vcs}, in the
narrow resonance limit we may relate the virial coefficients of the
two-component Fermi gas with those of a non-interacting Bose gas of
pairs at zero energy. Thus the thermodynamic variables at unitarity become those
of Eqs.~(\ref{eq:pbose}-c), where in this case $z^* = z$. 
We see that, as $R^*/\lambda$ increases, the thermodynamic
variables quickly approach those of the non-interacting Bose gas. This
further underlines how the narrow resonance limit is not strongly
interacting. We also see that the virial expansion appears to work
better in the narrow resonance limit, i.e., the second-order result becomes
closer to the fourth-order result. 
This is mainly because the odd orders of
the expansion  (e.g., third order) are suppressed as the system evolves into a
non-interacting gas of dimers.

\begin{figure}
\centering
\includegraphics[width=.859\linewidth]{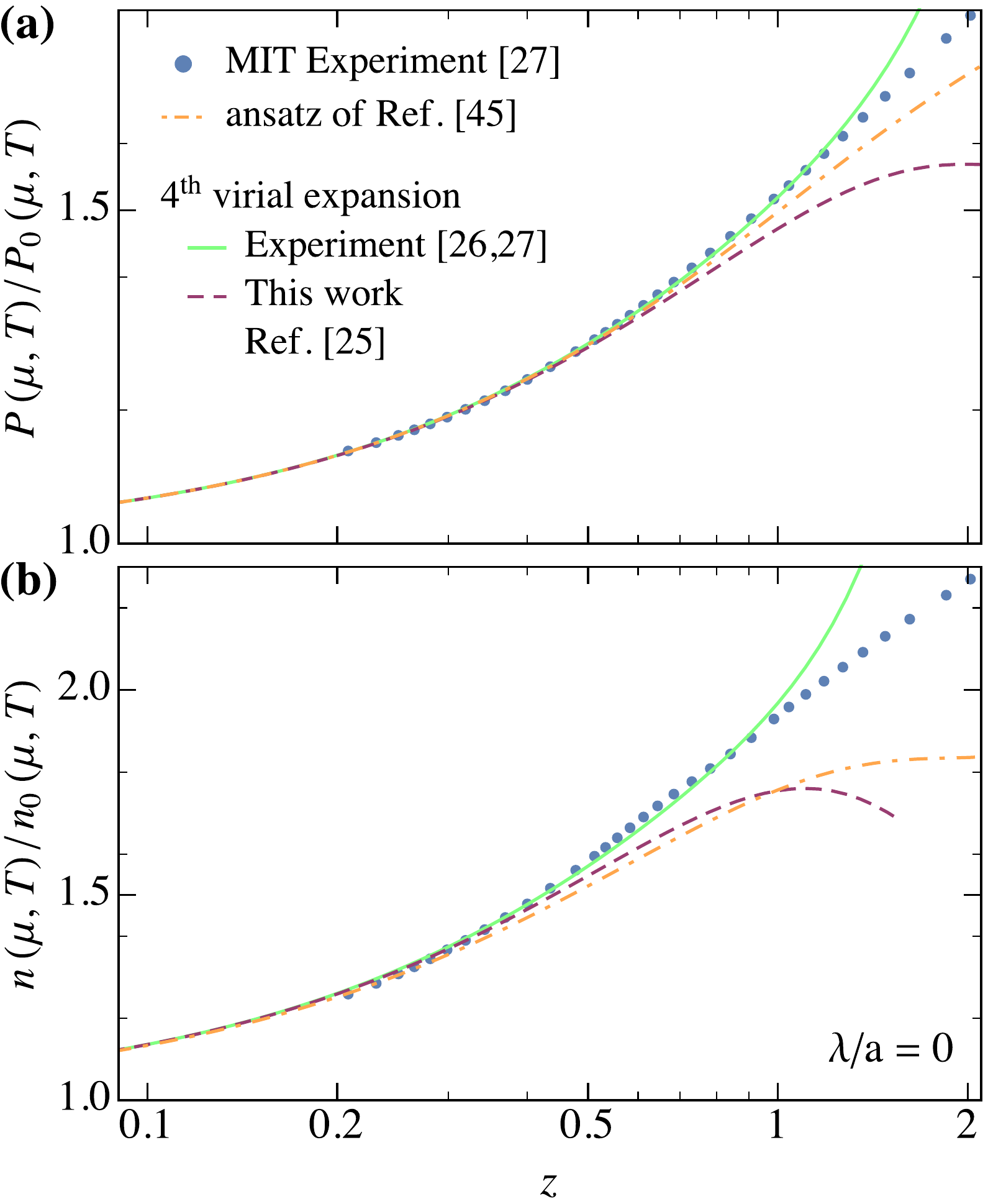}
\caption{Comparison between experiment \cite{Zwierlein2012} and
  various theories for the pressure and density at unitarity for a
  broad resonance. }
  \label{fig:MIT}
\end{figure}

In Fig.~\ref{fig:MIT}, we compare the virial expansion of the pressure
and density to the experimental data from the MIT experiment
\cite{Zwierlein2012}, which was conducted for a broad resonance in the
unitary limit, $1/a=0$. Although our approximation for the interaction
part of the fourth virial coefficient, 
$\Delta b_4^\text{Born}\approx0.06$, is smaller than the best fit to the
experimental data, $\Delta b_4^\text{exp.}\approx0.096$
\cite{Salomon2010,Zwierlein2012}, we observe a clear improvement when
comparing to the result of the previous theoretical estimate which had
a negative sign \cite{Rakshit2012}. We also apply the results of the
virial expansion ansatz developed in Ref.~\cite{Bhaduri2012} up to
20th order in the fugacity, and this approach is seen to work quite
well for the pressure and somewhat worse for the density. Finally, we
would like to point out that 
the fourth virial coefficient may even be slightly larger than the reported value of $0.096$, as this
would provide a 
better fit to the experimental data in the region around $z\sim0.5$, particularly in the case of the density. 
The fit as one approaches $z=1$ is less important since any deviation of the fourth-order virial expansion 
can be easily compensated by the inclusion of higher orders of the virial
expansion.

Thus far, we have worked in the grand canonical ensemble. This is
quite natural in ultracold atomic gases, where the chemical potential
across the external trap can typically be treated in the local density
approximation, $\mu(r)=\mu-V_\text{ext}(r)$. However, in a uniform
system it is often advantageous to consider the density instead of the
chemical potential. 
It is then convenient to work in the canonical ensemble, where the
relevant thermodynamic potential is the Helmholtz free energy
$F=F(n,T,V)$. In practice, one fixes the density and solves for the
fugacity using Eq.~\eqref{eq:densityvirial}, where the power series in
fugacity is truncated at a chosen order.  
When determing thermodynamic quantities $P$, $s$,
we only truncate
 the contribution arising from interactions and we treat the
ideal gas contribution exactly, i.e., the quantity
$\Delta\O\equiv\O-\O_0$ is expanded instead of the grand potential
$\O$.  
The advantage of this approach is that we recover the ideal Fermi gas 
to all orders in the Fermi limit. 
However, the different
truncation schemes should not lead to significantly different results
when the fugacity is small, which is a required condition for the
virial expansion.

\begin{figure}
\centering
\includegraphics[width=0.859\linewidth]{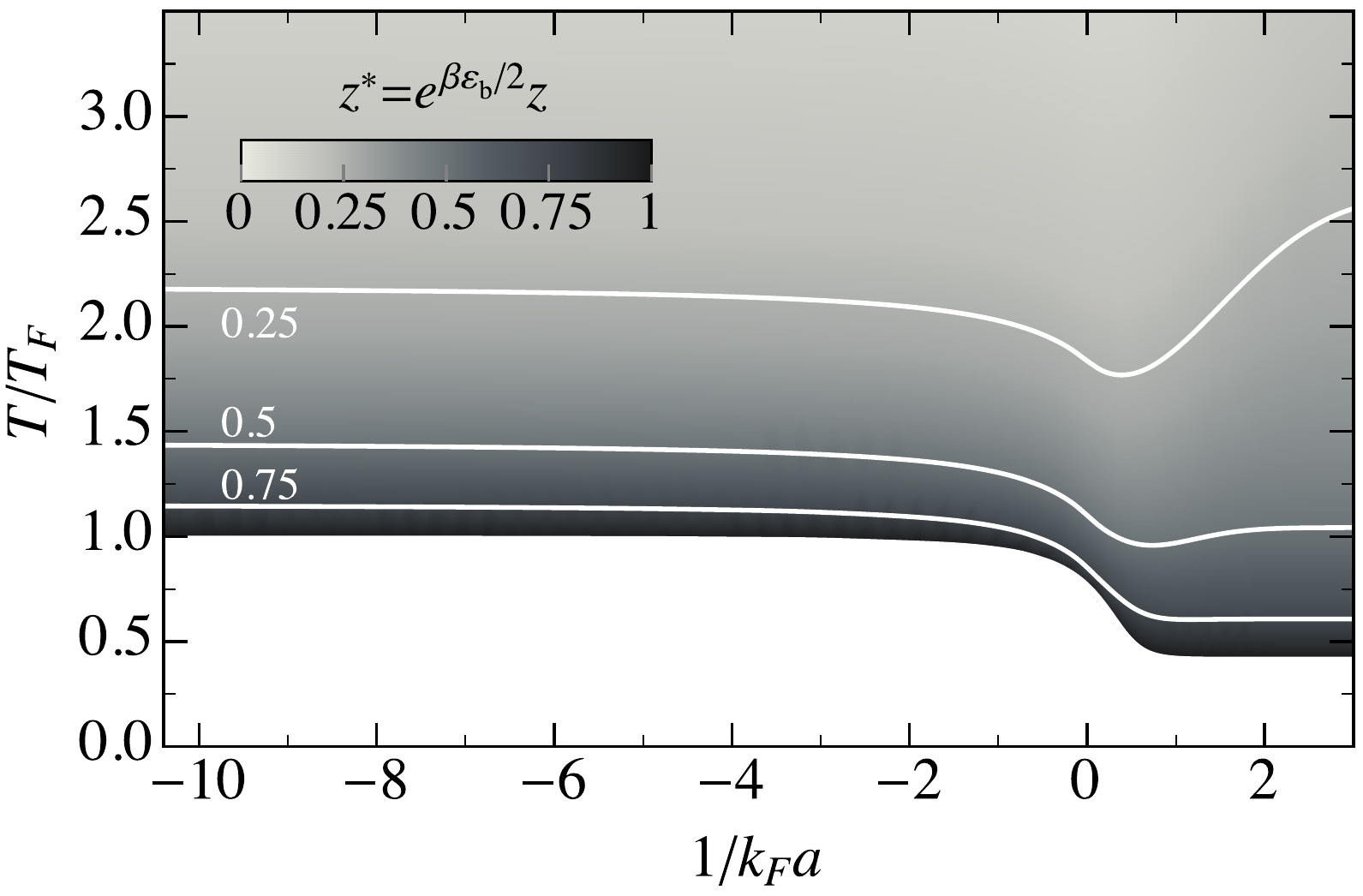}
\caption{Contour plot of the effective fugacity $z^*$ in the $T/T_F$
  versus $1/k_F a$ plane for a broad resonance with $R^* = 0$. The
  high-temperature expansion is considered up to the third virial
  coefficient, since our Born approximation for the fourth virial
  coefficient overestimates the effective dimer-dimer repulsion in the
  Bose regime. Note that we observe qualitatively similar
  behavior for a narrow resonance, and also if we limit ourselves to
  the virial expansion up to second order in the fugacity.}
\label{fig:fugacity-Virial2}
\end{figure}

To determine the region of validity of the virial expansion in this
parameter space, we calculate the fugacity as a function of
temperature $T/T_F$ and interaction strength $1/k_Fa$ using the 
virial expansion up to third order. 
Here, $T_F=\ef=\frac{k_F^2}{2m}$, with
$T_F$ and
$\ef$ the Fermi temperature and energy, 
respectively. In Fig.~\ref{fig:fugacity-Virial2}, we see that the
effective expansion parameter $z^*$, defined in Eq.~(\ref{eq:zstar}),
becomes comparable to unity for temperatures $T\sim T_F$, 
 as expected, 
and thus one should apply the virial expansion at a lower
temperature with some care. In addition, we note that at a given 
temperature $T/T_F$, $z^*$ has a minimum on the Bose side near unitarity,
corresponding to the point where the atomic and molecular states are
roughly equally accessible, and the density is spread between these states. 
Further into the Bose (Fermi) regimes, the density of dimers (atoms) builds up
and the system approaches quantum degeneracy.

\begin{figure*}
\centering
\includegraphics[width=.85\linewidth]{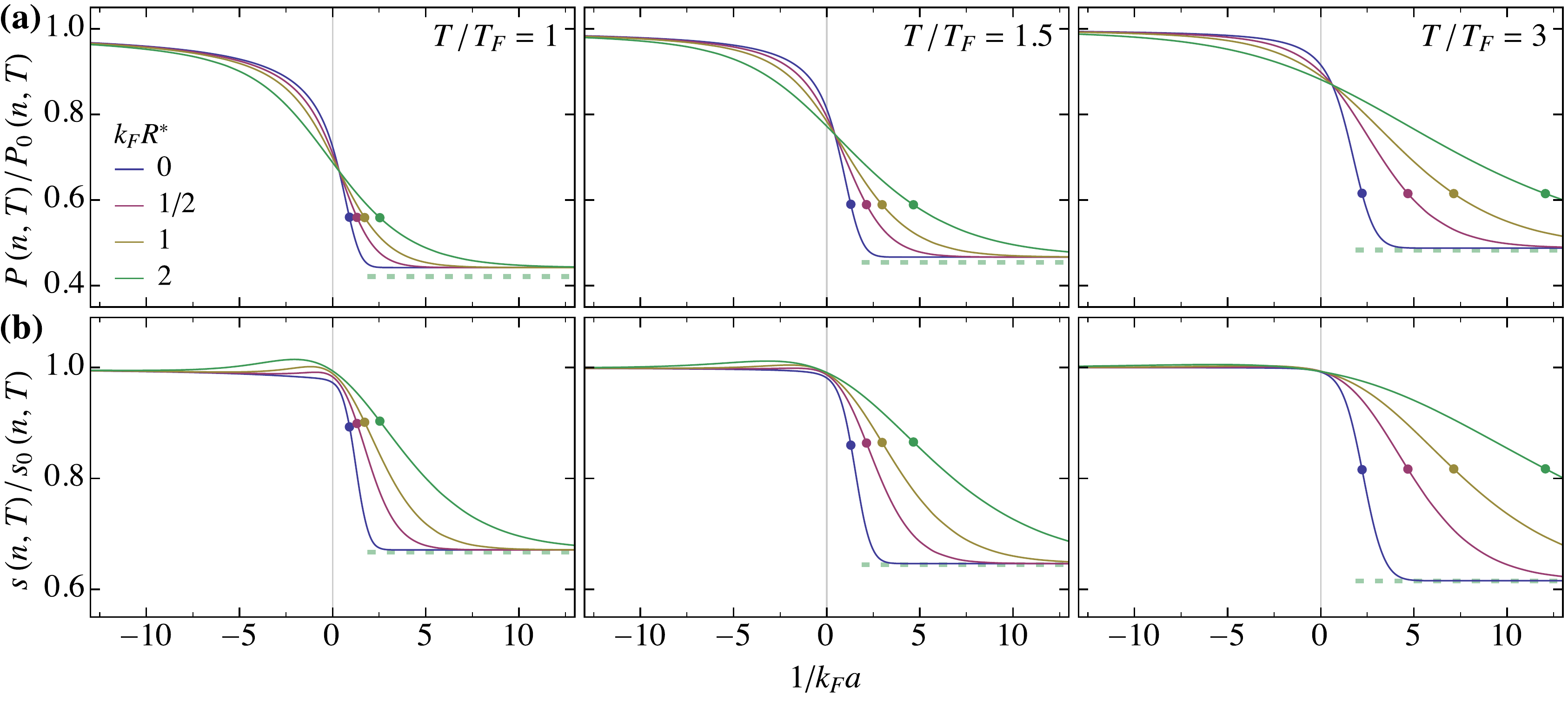}
\caption{The crossover behaviour 
of the pressure $P$ and entropy density $s$ 
  for both broad and narrow resonances.  
 The solid lines correspond to the results of the virial expansion where the non-interacting Fermi gas contribution 
 is treated exactly while the interacting part is only calculated up to the second virial coefficient.
  The solid circles mark the Fermi-Bose crossover point where $n\lambda^3 \approx \sqrt{2} e^{-\beta\eb}$.
  The dashed lines illustrate the thermodynamic values expected for an ideal Bose gas of dimers.
}
\label{fig:ThermodynamicCrossover}
\end{figure*}

Figure \ref{fig:ThermodynamicCrossover}(a) depicts the crossover
behaviour of the pressure and entropy density for various temperatures and range
parameters. We see how in the Fermi regime $1/k_Fa\lesssim-1$, these thermodynamic quantities
are close to those of the ideal Fermi gas, whereas in the opposite
regime they approach those of the ideal Bose gas.  Note that our results deviate from the expected value in the Bose limit
because the virial expansion we use only determines the interaction part up to second order in $z$ and therefore we only 
calculate the Bose gas up to first order in $z_{\rm Bose}$.
The crossover to tightly bound dimers occurs when the binding energy $\eb
\gg k_B T \log(\sqrt{2}/n\lambda^3)$, as discussed previously.
Interestingly, for a given $T/T_F$, this crossover point can be connected with particular values of $P/P_0$ and $s/s_0$, 
which appear to stay roughly constant as we vary $R^*$.

\section{Spectral function \label{sec:RF}}

Finally, we consider the spectral function $A(\k,\omega)=-2\Im
G(\k,\omega)$. This quantity is of great practical relevance, as it
may be probed in cold atom experiments by the use of radio-frequency
spectroscopy, allowing a direct investigation of many-body
effects. For instance, Ref.~\cite{chin2004} used this technique to
observe the formation of a pairing gap both above and below the
critical temperature for superfluidity. A typical experiment starts
from the strongly-interacting state, and applies a radio-frequency
pulse to transfer one of the $\up,\down$ spin states into an initially
unoccupied state. In the absence of interactions between the final and
initial spin states, the transfer rate is
\begin{equation}
  I(\k,\omega)\propto n_F(\omega)A(\k,\omega),\nn
\end{equation}
within linear response. Here, the Fermi distribution
$n_F(\omega)=(e^{\beta\omega}/z+1)^{-1}$ accounts for the probability
of the initial state being occupied. As usual, the Dyson equation
$G^{-1} = G_0^{-1} - \Sigma$ links the interacting with the
non-interacting single-particle Green's function. The self energy can
be expanded in powers of fugacity (see for instance
Refs.~\cite{Hu2010,Ngamp2013b}), and the lowest order contribution is
\begin{align}
&\Sigma(\k,\omega)=z
\sum_\q e^{-\beta\epsilon_{\q+\k}} 
T_2\left(\omega + \mu - \epsilon_{\k}+\frac{\epsilon_{\q}}{2}\right).
\end{align}

\begin{figure}õ
\centering
\includegraphics[width=0.95\linewidth]{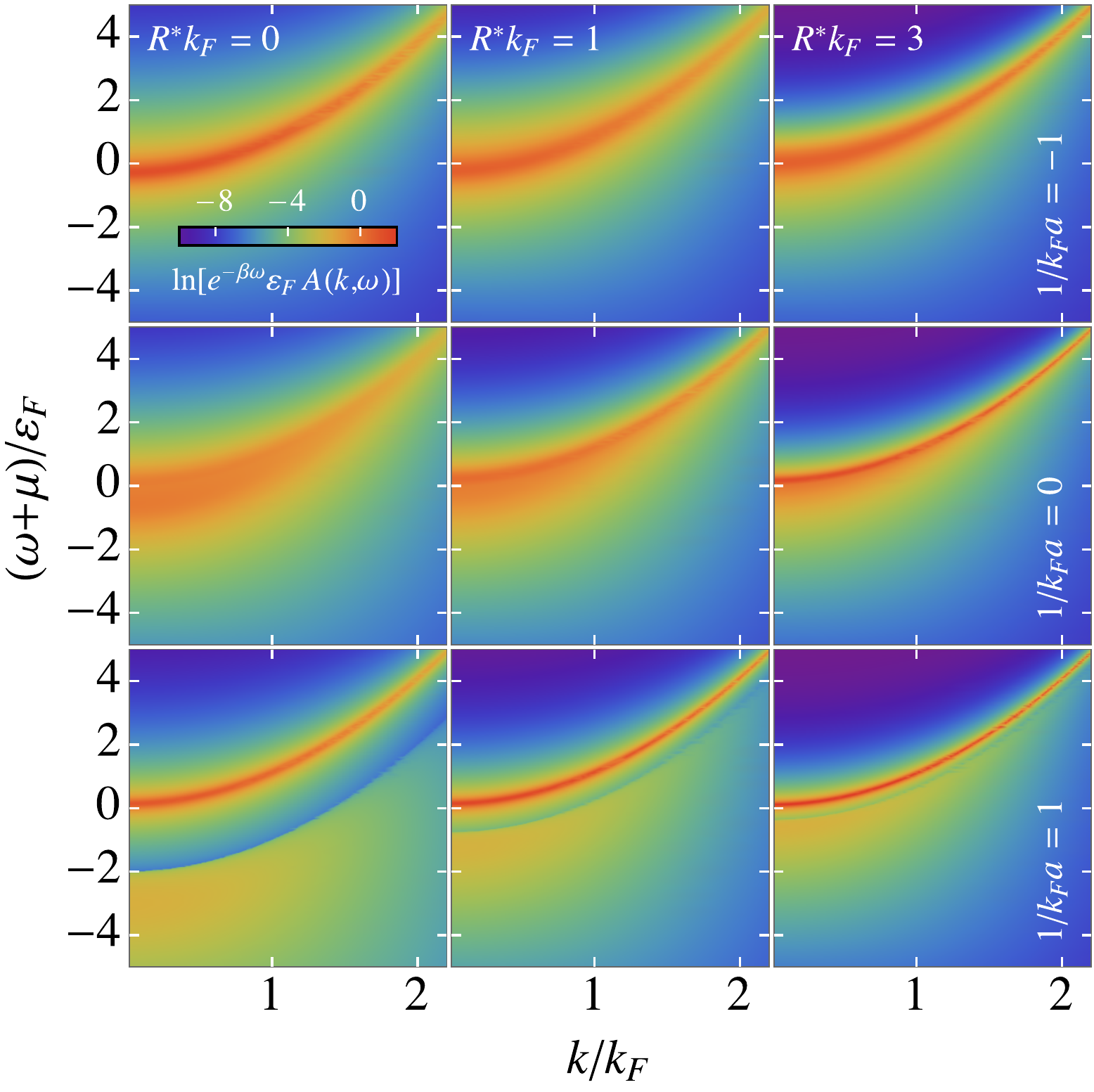}
\caption{ The occupied part of the spectral function for various
  values of the interaction parameter $1/k_Fa$ (top-to-bottom) and
  effective range $k_FR^*$ (left-to-right) near the crossover. They
  all correspond to the expansion parameter $ze^{\beta\eb/2} \approx
  0.2$. The short-dashed line corresponds to the free particle dispersion.}
\label{fig:b2spectra}
\end{figure}

The transition rate $I(\k,\omega)$ is illustrated in
Fig.~\ref{fig:b2spectra}. In the absence of interactions this would
simply correspond to a narrow peak at the free dispersion, indicated
by a dashed white line in the figures. However, interactions broaden the peak
and profoundly modify the spectrum: For positive scattering length, we
see that the spectral function develops a gap which separates the
lower band associated with paired atoms from the sharper upper peak
corresponding to unpaired atoms. Furthermore, the atomic peak is seen
to shift upward (downward) for sufficiently large positive (negative)
scattering lengths. We also consider the effect of a finite effective
range. Here, we clearly see that the atomic peak becomes sharper and
approaches the free dispersion, even at unitarity. This once again
indicates that the system approaches a non-interacting limit for
increasing $k_FR^*$.

\section{Conclusion \label{sec:conc}}

In this work, we have developed a diagrammatic formalism for computing
the coefficients of the virial expansion, and we have used it to elucidate
the high-temperature behavior of the resonant Fermi gas. We have focussed on
the crossover from the regime of unpaired atoms to the
regime of strongly bound pairs, and described the corresponding
limiting behaviors of the virial coefficients and thermodynamics. 
While the unitarity point is special in the sense that it is non-perturbative in $a$ 
and independent of an interaction length scale for a broad resonance, 
it is still possible to estimate values of the thermodynamic variables at this point since 
all virial coefficients are continuous functions of $\lambda/a$ and $R^*/\lambda$
throughout the crossover.
In particular, 
our approximate calculation of the fourth virial coefficient for a broad resonance 
is likely to be qualitatively very similar to the exact result, being an interpolation between known limits. 
Likewise, we believe our result of $\Delta b_4\approx0.06$ for the 
fourth virial coefficient at unitarity 
is close to the (currently unknown) exact value, 
and indeed it compares favorably with current experiment.
However, the strong non-monotonic behavior close to the resonance
also explains why it has been difficult to determine this coefficient accurately. 

We have also explored the behavior of the virial coefficients once a finite
effective range is introduced. Here we have shown that the
coefficients can be determined perturbatively, and we have calculated
the third and fourth virial coefficients for the first time.  We have argued that the narrow
resonance limit $R^*\gg\lambda$ corresponds to a non-interacting
mixture of closed-channel molecules and atoms. We have illustrated
this point by considering the behavior of the virial coefficients,
thermodynamic variables, and the spectral function. This should be
contrasted to the interaction energy which, for a resonant gas at zero
temperature, indeed increases with increasing $k_FR^*$
\cite{Ho2012}. However, this phenomenon is simply due to the depletion
of the Fermi sea as non-interacting pairs with zero energy are formed,
and we consider this to be a trivial pairing effect.

The formalism we have developed for the virial expansion is quite
general and is written in Section \ref{sec:virial} for a generic
short-range interaction in $d$ dimensions, for any mass ratio, and
for any spin imbalance. 
Thus, there are a number of
immediate extensions to our work on the resonant Fermi gas. For instance, 
one could consider the spin-imbalanced gas, where there are now two expansion parameters due to
the two different chemical potentials, $\mu_\up > \mu_\down$. In this case, the high-temperature
crossover should be considered as a crossover from a Fermi-Fermi
mixture with expansion parameters $e^{\beta\mu_\up}$ and
$e^{\beta\mu_\down}$, to a Bose-Fermi mixture described by effective
expansion parameters $e^{\beta(\mu_\up+\mu_\down+\eb)}$ and
$e^{\beta\mu_\up}$, assuming that $\mu_\up - \mu_\down$ is comparable to $\eb$ in this limit.

One can expect even richer behavior from the mass-imbalanced system
due to the possibility of additional scattering resonances. 
We have already described how the third and the fourth virial coefficients in the
Bose regime $\lambda/a\gg1$ may be related to the derivatives of the
low-energy atom-dimer and dimer-dimer scattering phase-shifts,
respectively. Whenever the few-body system is near-resonant, these
derivatives will change from 0 to $\pi/2$ over a small energy range
compared with the dimer binding energy, and consequently the virial
coefficients can be enhanced close to the formation of few-body bound
states, as discussed in Refs.~\cite{Pais1959,Bedaque2003}. In
particular, one expects an enhancement
of the $p$-wave contribution to the virial coefficient $B_{21}$ when
$m_\up/m_\down$ approaches $8.2$, the critical value 
for the formation of trimers in the $p$-wave channel when $a \gg R^*$
\cite{Kartavtsev2007}. Even below the critical mass ratio, such as in
a potassium-lithium mixture with mass ratio $40/6$, the atom-dimer
scattering is strongly enhanced, as discussed in
Refs.~\cite{Levinsen2009,Levinsen2011} and demonstrated in a recent
experiment \cite{Jag2014}. Likewise, tetramers are predicted to exist
in the $\up\up\up\down$ problem \cite{Levinsen2013,Blume2012} leading
to an enhancement of the coefficient $B_{31}$ close to the critical
mass ratio of $9.5$ close to unitarity 
\cite{Blume2012}. Indeed, the
sign-change of the third virial coefficient at unitarity as a function
of mass ratio \cite{Daily2012} 
results from  
the resonantly enhanced $p$-wave channel becoming dominant. 
Thus, few-body effects can profoundly impact the normal state of a heteronuclear
Fermi gas.

Finally, one may consider particles with other statistics. Thus far,
the virial expansion of the Bose gas has only been developed up to the
third virial coefficient at unitarity~\cite{Werner2013}, the results
showing an intriguing dependence on the three-body parameter
introduced by Efimov physics~\cite{Efimov}. It would thus be
interesting to consider the behavior across resonance, as in the
present work.


\begin{acknowledgments}
  We would like to thank Xavier Leyronas 
  for discussions, 
  and Martin Zwierlein for the data of Ref.~\cite{Zwierlein2012}.
  V.N. wishes to thank Aarhus Institute of Advanced
  Studies for hospitality during part of this work.
  M.M.P. acknowledges support from the EPSRC under Grant
No.\ EP/H00369X/2.
\end{acknowledgments}


\appendix


\input{stm.tex}


\input{BUhl_b3.tex}



\bibliography{virial}

\end{document}

%% file: stm.tex
\section{Three-body problem \label{app:STM}}

The three-body problem with short-range interactions is described
completely by the Skorniakov--Ter-Martirosian (STM) integral equation.
Originally introduced to calculate neutron-deuteron scattering
length~\cite{stm}, the diagrammatic representation of the STM equation
reads (see also Ref.~\cite{Bedaque1998,Levinsen2011})
\begin{align}
\begin{tikzpicture}[scale=0.8,baseline={([yshift=-.6ex]current  bounding  box.center)}]	
	\def	\radius{0.6}
	\draw	(-0.8*\radius,\radius) -- (0,\radius);
	\draw[T2]	(-0.8*\radius,0) -- (0,0);
	\draw	(\radius,\radius) -- (1.8*\radius,\radius);
	\draw[T2]	(\radius,0) -- (1.8*\radius,0);
	\draw[fill=gray!40]	(0,0) rectangle (\radius,\radius);
	\node	at	(0.5*\radius,0.5*\radius)	{$T_\text{ad}$};
\end{tikzpicture}\,
=\,
\begin{tikzpicture}[scale=0.8,baseline={([yshift=-.6ex]current  bounding  box.center)}]	
	\def	\radius{0.6}
	\draw	(\radius,\radius) -- (2*\radius,\radius) -- (1.8*\radius,0) -- (2.8*\radius,0);
	\draw[T2]	(\radius,0) -- (1.8*\radius,0);
	\draw[T2]	(2*\radius,\radius) -- (2.8*\radius,\radius);
\end{tikzpicture}
+
\begin{tikzpicture}[scale=0.8,baseline={([yshift=-.6ex]current  bounding  box.center)}]	
	\def	\radius{0.6}
	\draw	(-0.8*\radius,\radius) -- (0,\radius);
	\draw[T2]	(-0.8*\radius,0) -- (0,0);
	\draw	(\radius,\radius) -- (2*\radius,\radius)-- (1.8*\radius,0)-- (2.8*\radius,0);
	\draw[T2]	(\radius,0) -- (1.8*\radius,0);
	\draw[T2]	(1.8*\radius,\radius) -- (2.8*\radius,\radius);
	\draw[fill=gray!40]	(0,0) rectangle (\radius,\radius);
	\node	at	(0.5*\radius,0.5*\radius)	{$T_\text{ad}$};
\end{tikzpicture}\,.
\label{STMdiag}
\end{align}
Here $T_\text{ad}$ is the atom-dimer scattering $T$ matrix; the solid
lines denote free atom propagators and the dotted lines the two-body
$T$ matrix. We then write Eq.~(\ref{STMdiag}) for an $\up\up\down$
system as an integral equation
\begin{widetext}
\begin{align}
T_\text{ad}(\mathbf k ,k_0;\mathbf p, p_0;E) 
= \zeta  \Bigg[&
  g^2ZG_\downarrow(\mathbf{-k-p},E-k_0-p_0)
\nn\\&
- \int\frac{d^3\q}{(2\pi)^3}  \int 
  \frac{d q_0}{2\pi i}
G_\uparrow(\mathbf q,q_0)
G_\downarrow(\mathbf{-p-q},E-p_0-q_0)
T_2(\mathbf{-q},E-q_0)T_\text{ad}(\mathbf k ,k_0;\mathbf q, q_0;E)\Bigg].
\label{STM}
\end{align}
\end{widetext}
We let the incoming [outgoing] atom and dimer have four-momenta
$(\k,\ek)$ and $(-\k,E-\ek)$ [$(\p,\ep)$ and $(-\p,E-\ep)$],
respectively.  The free atom propagators is given by
$G_{\uparrow,\downarrow}(\mathbf k,k_0) =
(k_0-k^2/2m_{\up,\down}+i0)^{-1} $ and $T_2(\q,q_0)\equiv
T_2(q_0+q^2/2(m_\up+m_\down)+i0)$ is the two-body $T$ matrix.  The
residue at the pole of the two-body $T$ matrix --- $\displaystyle g^2Z
\equiv \lim_{s\to 0}s\times T_2(-\eb+s) $ --- is included for correct
normalization of the external dimer propagators.  The factor $\zeta$
in Eq.~(\ref{STM}) describes the quantum statistics and is equal to
$-1$ for the fermionic case, $+1$ for heteronuclear bosons, and $+2$
for identical bosons.

The integration over $q_0$ can be done by closing the contour in the
lower half plane. This encloses one pole coming from
$G_\uparrow(\mathbf q,q_0)$ at $q_0=\eq$. Next, we take the partial
wave projection and use the on-shell conditions: $k_0 = \ek$ and $p_0
= \ep$. The equation~(\ref{STM}) becomes
\begin{align}
&T_{\text{ad},\ell}(k,p;E) =\zeta \Bigg[
  g^2Z g_\ell(k,p;E)
\nn \\&+ \sum_q
g_\ell(p,q;E)
T_2(E-q^2/2m_\text{21})T_{\text{ad},\ell}(k,q;E)\Bigg].
\end{align}
where $m_\text{21}^{-1} = m_\up^{-1}+(m_\up+m_\down)^{-1}$.

The partial wave projection is defined as follows (energy omitted)
\begin{align}
T_{\text{ad},\ell}(p,q) &= 
 \int_{-1}^{1}\frac{dx}{2}P_\ell(x) T_\text{ad}(\p,\q), \nn
\\ T_\text{ad}(\p,\q) &= 
\sum_{\ell\ge0} (2\ell+1) P_\ell(x) T_{\text{ad},\ell}(p,q), 
\end{align}
where $x=\cos(\hat \p\cdot\hat \q)$ and $P_\ell(x)$ denote the
Legendre polynomials. Similarly, one may write down the partial-wave
projected $\down$ propagator,
\begin{align}
  g_\ell(p,q) &= \int_{-1}^{1}\frac{dx}{2}P_\ell(x)
  G_\down(\p+\q,E-\ep-\eq)\nn
  \\
  &=\frac{m_\down}{pq}Q_\ell\left[\frac{m_\down}{pq}\left(E -
      \frac{p^2+q^2}{2m_r}\right)\right],
\end{align}
where $Q_\ell(z) = \int_{-1}^1 (z-x)^{-1}P_\ell(x) \,d x/2$ is the Legendre
function of the second kind.

In general, there might not exist a two-body bound state, e.g., when
$a<0$ in 3D. It is thus useful to define the three-body $T$ matrix
$t_3\equiv T_\text{ad}/g^2Z$ such that the external dimer propagators
are removed from Eq.~\eqref{STMdiag}.  Therefore, we have
\begin{align}
&t_{3,\ell}(k,p;E) =\zeta \Bigg[
  g_\ell(k,p;E)
\nn\\&  + \sum_q g_\ell(p,q;E)T_2(E-q^2/2m_\text{21})t_{3,\ell}(k,q;E)\Bigg].
\end{align}
%

%% file: BUhl_b3.tex
\section{Beth-Uhlenbeck expression for $\Delta b_3$ in the Bose
  regime \label{app:BU3body}}

In the Bose limit, the three-body problem reduces to the much simpler
atom-dimer scattering problem. 
This is effectively a two-body problem and thus one may readily modify 
the Beth-Uhlenbeck formula, see Eq.~(\ref{eq:B-Uh}), to obtain the 
third virial coefficient. 
Indeed, we have shown that the dominant diagram in this limit 
is the one describing atom-dimer scattering -- see, 
Eq.~(\ref{b3scaling:atomdimer}). 
That is, we have
%
\begin{align}
\nn
B_{21} &\overset{\lambda\gg a}{\approx}
\frac{\lambda^3}{V}\oint'_E
\begin{tikzpicture}[baseline={([yshift=-.6ex]current  bounding  box.center)}]
\draw[blue] (0.2,-0.1) -- (0,-0.1);
\draw[red] (0,0.1) -- (0.2,-0.1);
\draw[T2] (0.2,-0.1) -- (0.4,-0.1); \draw[blue] (0,0.3) -- (0.4,0.3);
\fill[gray,opacity=0.5] (0.4,-0.1) rectangle (0.8,0.3) ;
\node at (0.6,0.1) {$t_3$};
\draw[blue] (0.8,-0.1) -- (1.2,-0.1) ;
\draw[T2] (0.8,0.3) -- (1,0.3);
\draw[blue] (1.2,0.3) -- (1,0.3);\draw[red] (1,0.3) -- (1.2,0.1);
\node[blue] at (-0.1,-0.1) {1};\node at (-0.1,0.1) [red]{1};\node[blue] at (-0.1,0.3) {2};
\node[blue] at (1.3,-0.1) {2};\node at (1.3,0.1) [red]{1};\node[blue] at (1.3,0.3) {1};
\end{tikzpicture}
\\
&\overset{\phantom{\lambda\gg a}}{=}
\left(\frac{M_{21}}{2m_r}\right)^{\frac{d}{2}}\oint'_E\sum_\p
t_3(\p,\p;E)\nn\\&\hspace{1.5cm}\times
T_2^2\left(E-\frac{p^2}{2m_{21}}\right)
\frac{d}{dE}T_2^{-1}\left(E-\frac{p^2}{2m_{21}}\right)
\nn.
\end{align}
%

As $E\rightarrow-\eb$, the `free' atom-dimer propagator is given by
\begin{align}
\bra\k \hat G_{\text{ad}}(E)\ket\k \approx 
\frac{T_2\left(E-\frac{k^2}{2m_{21}}\right)}{g^2Z} 
\approx \frac{1}{E_\text{col}-\frac{k^2}{2m_{21}}},
\end{align}
where $\pm\k$ are the momenta of the atom and dimer in their
center-of-mass frame, the collision energy is given by $E_\text{col} =
E+\eb$ and $g^2Z$ is the residue at $T_2$'s pole given in
Eq.~(\ref{eq:residue}).  In addition, we note that the atom-dimer
forward scattering $T$ matrix is given by
\begin{align}
T_\text{ad}(k^2/2m_{21}) = g^2Zt_3(\k,\k;k^2/2m_{21}-\eb).
\label{eq:Tforward}
\end{align}
Therefore, we obtain
\begin{align}
B_{21} &\overset{\lambda\gg a}{\approx}
\left(\frac{M_{21}}{2m_r}\right)^{\frac{d}{2}}
\oint'_E\sum_{\p,\ell}
\frac{\xi_\ell T_{\text{ad},\ell}(E_\text{col})}{\left(E_\text{col}-\frac{p^2}{2m_{21}}\right)^2}
\nn\\
&\overset{\phantom{\lambda\gg a}}{\approx}
 \left(\frac{M_{21}}{2m_r}\right)^{\frac{d}{2}}
e^{\beta\eb} \oint'_E\sum_{\p,\ell}
\frac{\xi_\ell T_{\text{ad},\ell}(E)}{\left(E-\frac{p^2}{2m_{21}}\right)^2},
\label{B21_preBUhl}
\end{align}
where we used $\frac{d}{dE} T_2^{-1} \approx g^2Z$
and in the last step, the energy origin is shifted to $-\eb$. 
The full scattering $T$ matrix is given by the 
partial wave sum 
$T_\text{ad}(s) = \sum_\ell \xi_\ell T_{\text{ad},\ell}(s)$, 
where we have used the fact that only forward atom-dimer scattering is
present in Eq.~\eqref{eq:Tforward}.

As shown in Section \ref{sect:B11}, the sum over $\p$ is in fact the sum over
the two-body spectrum which is equal to the derivative of the inverse two-body
$T$ matrix. That is, we have
\begin{align}
\sum_{\p}\left(E-\frac{p^2}{2m_{21}}\right)^{-2}
\equiv
\frac{d}{dE}\sum_\ell\xi_\ell T^{-1}_{\text{ad},\ell}(E).
\end{align}
Following the same analysis, we recast 
Eq.~(\ref{B21_preBUhl}) in the Beth-Uhlenbeck form 
\begin{align}
B_{21}\overset{\lambda\gg a}{\approx} \left(\frac{M_{21}}{2m_r}\right)^{\frac{d}{2}}
e^{\beta\eb} \sum_{\ell}\xi_\ell\int_0^\infty\frac{dk}{\pi}e^{-\beta \frac{k^2}{2m_{21}}}
\delta'_{\text{ad},\ell}(k),
\end{align}
which concludes this Appendix.